\documentclass[preprint]{elsarticle}

\usepackage{mathtools}
\usepackage{algorithmicx}
\usepackage{algpseudocode}
\usepackage{algorithm}
\usepackage{amsfonts}
\usepackage{amssymb}
\usepackage{subcaption}
\usepackage{graphicx,import,color}
\usepackage{soul}
\sethlcolor{white}
\newcommand{\hll}[1]{\colorbox{white}{$\displaystyle #1$}}

\usepackage{lineno,hyperref}

\newdefinition{rmk}{Remark}

\journal{Computer Methods in Applied Mechanics and Engineering}

\begin{document}

\begin{frontmatter}

\title{An efficient and robust monolithic approach to phase-field quasi-static brittle fracture using a modified Newton method}

%% or include affiliations in footnotes:
\author[mymainaddress]{Olivier Lampron}

\author[mymainaddress]{Daniel Therriault}

\author[mymainaddress]{Martin Lévesque\corref{mycorrespondingauthor}}
\cortext[mycorrespondingauthor]{Corresponding author}
\ead{martin.levesque@polymtl.ca}

\address[mymainaddress]{Laboratory for Multiscale Mechanics, Department of Mechanical Engineering, Polytechnique Montreal, 2900 boul. Edouard-Montpetit, Montreal, QC, Canada}

\begin{abstract}
Variational phase-field methods have been shown powerful for the modeling of complex crack propagation without a priori knowledge of the crack path or ad hoc criteria. However, phase-field models suffer from their energy functional being non-linear and non-convex, while requiring a very fine mesh to capture the damage gradient. This implies a high computational cost, limiting concrete engineering applications of the method. In this work, we propose an efficient and robust fully monolithic solver for phase-field fracture using a modified Newton method with inertia correction and an energy line-search. To illustrate the gains in efficiency obtained with our approach, we compare it to two popular methods for phase-field fracture, namely the alternating minimization and the quasi-monolithic schemes. To facilitate the evaluation of the time step dependent quasi-monolithic scheme, we couple the latter with an extrapolation correction loop controlled by a damage-based \hl{criterion}. Finally, we show through four benchmark tests that the modified Newton method we propose is straightforward, robust, and leads to identical solutions, while offering a reduction in computation time by factors of up to 12 and 6 when compared to the alternating minimization and quasi-monolithic schemes.

\end{abstract}


\end{frontmatter}

%\linenumbers

\section{Introduction} \label{Introduction}

The variational reformulation of Griffith’s theory~\cite{griffith_phenomena_1921} proposed by Francfort and Marigo~\cite{francfort_revisiting_1998} and followed by Bourdin et al.'s phase-field regularization~\cite{bourdin_numerical_2000} offered a robust mathematical and computational framework for the modeling of quasi-static brittle fracture. Based on a smooth representation of the discontinuities through a damage field, variational phase-field models compute the nucleation and propagation of complex crack paths through the minimization of an energy functional. As opposed to discrete approaches dealing explicitly with cracks like Linear Elastic Fracture Mechanics (LEFM) or Cohesive Zone Models (CZMs), the phase-field method for fracture can model crack propagation without the need for remeshing, a priori knowledge of the crack path or ad hoc criteria~\cite{wu_phase-field_2019}. Owing to its elegant simplicity and its compatibility with classical finite element methods, the phase-field method has been applied and extended to hydraulic fracture~\cite{mikelic_phase-field_2015,wilson_phase-field_2016}, ductile fracture~\cite{miehe_phase_2016,borden_phase-field_2016}, stress corrosion cracking~\cite{nguyen_phase_2017,martinez-paneda_phase_2018} and brittle fracture in heterogeneous structures~\cite{hansen-dorr_phase-field_2019,nguyen_phase_2019}.

A well-known downside of phase-field models is their high computational cost. The first explanation for this situation is the necessity of a very fine mesh for the finite element method to accurately capture the strong damage gradient in the fracture zone, yielding large linear systems to be solved. Adaptive mesh techniques~\cite{heister_primal-dual_2015,ferro_anisotropic_2018,mang_mesh_2020} and multiscale methods~\cite{patil_adaptive_2018, nguyen_multiscale_2019} have been developed and appear to successfully reduce the number of degrees of freedom to a minimum. 

The second explanation for the high computational cost is the non-linear and non-convex nature of the underlying energy functional of phase-field models~\cite{bourdin_numerical_2000}. For non-convex problems, the conventional Newton's method is inadequate since the Hessian can become indefinite and hinder the convergence to a critical point. The first method proposed in the literature to solve phase-field models was the alternating minimization algorithm~\cite{bourdin_numerical_2000}. Stemming from the observation that the functional is convex in each variable when the other field is fixed, the alternating minimization scheme alternatively solves the kinematic and damage sub-problems until convergence. Widely adopted by the community~\hl{\mbox{\cite{martinez-paneda_phase_2018,hansen-dorr_phase-field_2019,miehe_phase_2010,amor_regularized_2009,borden_phase-field_2012,ambati_phase-field_2015,nguyen_phase-field_2016,marigo_overview_2016,pham_experimental_2017,tanne_crack_2018,gerasimov_penalization_2019}}}, the algorithm is robust, but exhibits a slow convergence rate due to the decoupling of the sub-problems~\cite{gerasimov_line_2016,farrell_linear_2017,wu_bfgs_2020}. Significant gains in efficiency were obtained using a combination of preconditioned alternating minimization scheme and classical Newton's method~\cite{farrell_linear_2017}\hl{, a staggered scheme relying on a truncated modified Newton method~\mbox{\cite{kirkesaether_brun_iterative_2020}}}, or a semi-implicit form of the staggered scheme~\cite{lu_efficient_2020}. \hl{An accelerated staggered scheme combining Anderson acceleration and over-relaxation was also shown to yield significant decrease in computation time in~\mbox{\cite{storvik_accelerated_2021}}.} Nevertheless, such methods require either the tuning of \hl{numerical parameters} or a time step convergence study, implying additional computation, while still suffering from the decoupling between the displacement and damage fields.

\hl{The quasi-monolithic approach~\mbox{\cite{heister_primal-dual_2015}} is a non-partitioned and robust alternative to the staggered scheme relying on the linearization of the kinematic sub-problem through a time extrapolation of the phase-field~\mbox{\cite{wick_multiphysics_2020}}. This partial linearization allowing to solve simultaneously both sub-problems was used in~\mbox{\cite{lee_phase-field_2018,jodlbauer_matrix-free_2019,mang_phase-field_2020}}}. The quasi-monolithic scheme was also reported efficient and robust in~\cite{wick_error-oriented_2017}, but the relaxation of the applied forces and a delayed crack growth was observed. Furthermore, the delayed and relaxed evolution seems to be amplified under unstable crack growth, where the damage field evolves rapidly. This suggests that the quasi-monolithic scheme generates a numerical viscosity, requiring smaller time steps and, therefore, an increased computational cost to accurately capture rapidly growing cracks. To the best of our knowledge, the efficiency of the alternating minimization algorithm and the quasi-monolithic scheme has never been compared.

\hl{A fully monolithic scheme would have the potential to be more efficient than the alternating and quasi-monolithic methods since it would preserve the strong coupling between the two fields.} As reported by Wick~\cite{wick_modified_2017}, obtaining a monolithic scheme for phase-field models for fracture is a difficult task. Numerous monolithic algorithms have been proposed, such as a Newton method using a modified line-search algorithm~\cite{gerasimov_line_2016}, an error-oriented Newton’s method~\cite{wick_error-oriented_2017}, a modified Newton method with Jacobian modification~\cite{wick_modified_2017}, \hl{and a multilevel trust region method~\mbox{\cite{kopanicakova_recursive_2020}}}. Probably due to their complexity, variable robustness, or problem-dependent efficiency, these monolithic methods have yet to impose themselves, with the alternating minimization algorithm and quasi-monolithic scheme still being vastly more represented in the literature.

More recently, Wambacq et al.~\cite{wambacq_interior-point_2021} proposed a monolithic solver relying on a modified Newton method with inertia correction inspired by~\cite{wachter_implementation_2006}. Although the modified Newton method showed interesting performance, their work focused on the efficiency of the interior-point method to enforce the irreversibility condition present in phase-field models for fracture. Consequently, the efficiency of the monolithic solver was only compared to the alternating algorithm when coupled with the interior-point method. The performances of the two solvers were concluded to be problem-dependent in this case.

We also point out that few popular formulations, such as the history variable approach of Miehe and co-workers~\cite{miehe_phase_2010}, or the hybrid model from~\cite{ambati_review_2015}, although computationally efficient, have the downfall of being variationally inconsistent. Gerasimov and De Lorenzis came to the conclusion that the approach based on the history variable proposed by Miehe~\cite{miehe_phase_2010} leads to the solution of a problem that is not equivalent to the variational one~\cite{gerasimov_penalization_2019}. \hl{Therefore, we focus in this work on the efficiency of solvers for phase-field models that do not rely on the history variable.}

In this work, we formulate a fully monolithic and implicit solver relying on a modified Newton method with inertia correction of the Jacobian and a backtracking energy line-search, similar to~\cite{wambacq_interior-point_2021,wachter_implementation_2006}. We then compare the performance of the alternating minimization algorithm, the quasi-monolithic scheme and the modified Newton solver on four benchmark tests selected from the phase-field literature. To eliminate the time step dependency of the quasi-monolithic scheme and facilitate its benchmarking, we couple the latter with an extrapolation correction loop controlled by a damage evolution \hl{criterion}. The benchmarking demonstrates that not only is the modified Newton method a robust monolithic solver for variational phase-field models, but it is also significantly more efficient than the quasi-monolithic and alternating minimization algorithms.

The paper is organized as follows. In Section~\ref{Background}, we recall the formulation of variational phase-field models, describe the adopted treatment of the irreversibility constraint, and write the Euler-Lagrange equations of the minimization problem along with their discretization. We also present the alternating minimization algorithm and the quasi-monolithic scheme. In Section~\ref{New solution algorithms}, we present a monolithic solver for variational phase-field models based on a modified Newton method and describe the quasi-monolithic scheme augmented with an extrapolation correction loop. Section~\ref{Numerical results} compares the performance of the three studied algorithms on four different benchmark tests and discusses the results. Finally, our conclusions are presented in Section~\ref{Conclusions}.

\section{Background} \label{Background}

\subsection{Variational modeling of brittle fracture} 
Here we define the phase-field model for brittle fracture, including the strain energy decomposition and treatment of the irreversibility condition adopted, on which is performed the benchmarking of the three solvers. Then, the Euler-Lagrange equations of the unconstrained variational problem and their spatial discretization are defined.

\subsubsection{Variational formulation}
Let $\Omega \subset {\rm I\!R}^\delta$ ($\delta = 2,3$) be the domain representing a crack-free brittle elastic body. The boundary $\partial \Omega$ is subject to time-dependent Dirichlet and Neumann conditions, $\Gamma_D \subset \partial \Omega$, $\Gamma_N \subset \partial \Omega$, with $\Gamma_D \cup \Gamma_N = \partial \Omega$ and $\Gamma_D \cap \Gamma_N = \emptyset$. We also introduce an evolving crack surface $\Gamma_c \subset \Omega$. We define the displacement function $\boldsymbol{u}:\Omega \setminus \Gamma_c \subset {\rm I\!R}^\delta \rightarrow {\rm I\!R}^\delta$ and the linearized elastic strain $\boldsymbol{e}(\boldsymbol{u})\coloneqq \frac{1}{2}(\nabla \boldsymbol{u}^T + \nabla \boldsymbol{u})$. We can therefore define the elastic potential energy $\Psi(\boldsymbol{u}) = \frac{1}{2} \boldsymbol{\sigma}(\boldsymbol{u}):\boldsymbol{e}(\boldsymbol{u})$, where $\boldsymbol{\sigma}(\boldsymbol{u})$ is the stress tensor. Finally, we denote the critical energy release rate of the brittle material with $G_c$.

Neglecting the external forces, Francfort and Marigo's \hl{reformulation} of quasi-static brittle fracture~\cite{francfort_revisiting_1998} leads to the following variational problem
\begin{equation}
    \boldsymbol{u},\Gamma_c = \textrm{arg}\,\textrm{min } \mathcal{E}(\boldsymbol{u},\Gamma_c ) \quad \textrm{s.t.} \quad \dot{\Gamma}_c \geq 0,
\end{equation}
with the energy functional
\begin{equation}
	\mathcal{E}(\boldsymbol{u}, \Gamma_c) = \int_{\Omega \setminus \Gamma_c} \Psi(\boldsymbol{u}) \, \text{d}\boldsymbol{x} + \int_{\Gamma_c} G_c \, \text{d}s.
\end{equation}

Phase-field damage models use the Ambrosio and Tortorelli elliptical regularization~\cite{ambrosio_approximation_1990} to overcome problems associated with the numerical treatment of the discontinuous crack $\Gamma_c$, as proposed initially in~\cite{bourdin_numerical_2000}. The regularization replaces the sharp crack by a diffuse damage band through the phase-field variable, denoted by $d:\Omega \rightarrow [0,1]$, where $d=0$ describes the undamaged material and $d=1$ describes the fully damaged material. The displacement function becomes $\boldsymbol{u}:\Omega \subset {\rm I\!R}^\delta \rightarrow {\rm I\!R}^\delta$. The different phase-field models for brittle fracture are characterized mainly by the type of crack function $\alpha(d)$ and degradation function $g(d)$ adopted. They can generally be expressed as

\begin{equation}
    \mathcal{E}_l(\boldsymbol{u},d) = \int_{\Omega} g(d)\Psi(\boldsymbol{u}) \, \text{d}\boldsymbol{x} + \frac{G_c}{c_\alpha}\int_{\Omega}\Big(\frac{\alpha(d)}{l}+l|\nabla d|^2 \Big)\text{d}\boldsymbol{x},
\end{equation}
with the normalization parameter $c_\alpha \coloneqq 4\int_0^1 \sqrt{\alpha(z)} \, \text{d}z$ \cite{tanne_crack_2018,gerasimov_penalization_2019}. The regularization $\alpha(d)$ and the characteristic length $l$ define the shape and width of the smeared crack while the degradation function $g(d)$ describes the dependency of the apparent stiffness on the damage field. In this paper, we focus on the so-called $AT_1$ model with its linear crack function $\alpha(d) = d$ and quadratic degradation function $g(d)=(1-d)^2$ introduced by Pham et al.~\cite{pham_gradient_2011}, yielding the functional
\begin{equation}
\label{AT1}
    \mathcal{E}_l(\boldsymbol{u},d) = \int_{\Omega} (1-d)^2\Psi(\boldsymbol{u}) \, \text{d}\boldsymbol{x} + \frac{3G_c}{8}\int_{\Omega}\Big(\frac{d}{l}+l|\nabla d|^2 \Big)\text{d}\boldsymbol{x}.
\end{equation}
With the regularized energy functional relying on the phase-field variable, the variational problem now reads
\begin{equation}
\label{VP}
    \boldsymbol{u},d = \textrm{arg}\,\textrm{min } \mathcal{E}_l(\boldsymbol{u},d ) \quad \textrm{s.t.} \quad 0 \leq d \leq 1 \quad \textrm{and} \quad \dot{d} \geq 0.
\end{equation}

As opposed to the widely used $AT_2$ model, the $AT_1$ model prevents damage initiation at low strain-energy density, therefore exhibiting an elastic phase. Furthermore, the linear crack function of the $AT_1$ model leads to a finite localization band, implying the convergence to the Griffith energy without the necessity of the characteristic length $l$ to approach zero~\cite{tanne_crack_2018}. However, the constraint $d \geq 0$ from the minimization problem \eqref{VP} is no longer naturally respected.

A well-known issue with formulation \eqref{AT1} is its inability to distinguish between energy generated from tensile or compressive loading. Consequently, interpenetration of crack surfaces and non-physical cracks were observed early in the developments of the variational approach to fracture~\cite{bourdin_numerical_2000}. Multiple methods using energy splits have been proposed to elude this difficulty~\cite{lancioni_variational_2009, wu_novel_2018, bilgen_crack-driving_2019,dijk_strain_2020}, such as the volumetric-deviatoric split~\cite{amor_regularized_2009} adopted in~\cite{schluter_phase_2014,wu_phase-field_2017} and the spectral split~\cite{miehe_phase_2010} adopted in~\cite{pham_experimental_2017,wick_modified_2017}. In this work, we adopt the spectral decomposition of the strain tensor in its positive and negative components, as proposed by Miehe~\cite{miehe_phase_2010}, with
\begin{equation}
	\boldsymbol{e}(\boldsymbol{u}) = \boldsymbol{e}^+(\boldsymbol{u}) + \boldsymbol{e}^-(\boldsymbol{u}),
\end{equation}
and
\begin{equation}
	\boldsymbol{e}^+(\boldsymbol{u}) = \boldsymbol{S}(\boldsymbol{u}) \boldsymbol{\Lambda}^+(\boldsymbol{u}) \boldsymbol{S}^T(\boldsymbol{u}),
\end{equation}
where $\boldsymbol{\Lambda}^+$ is the diagonal matrix containing the positive eigenvalues $\lambda_i^+$, and $\boldsymbol{S}$ is the matrix containing the corresponding eigenvectors $\boldsymbol{s}_i$~\cite{wick_error-oriented_2017}. Dropping the dependencies in $\boldsymbol{u}$, the positive and negative stress of an isotropic material now read
\begin{equation}
\boldsymbol{\sigma}^+ = \lambda \langle \text{tr}(\boldsymbol{e}) \rangle_{+} \boldsymbol{I} + 2\mu \boldsymbol{e}^+,
\end{equation}
\begin{equation}
\boldsymbol{\sigma}^- = \lambda \langle \text{tr}(\boldsymbol{e}) \rangle_{-} \boldsymbol{I} + 2\mu \boldsymbol{e}^-,
\end{equation}
where $\boldsymbol{I}$ is the second order identity tensor and ($\lambda$, $\mu$) are the Lamé coefficients. $\langle z \rangle_{\pm}$ designates the positive or negative ramp function of $z$, where $\langle z \rangle_{+} = \text{max}(0,z)$ and $\langle z \rangle_{-} = \text{min}(0,z)$. The energy functional using the energy split becomes
\begin{equation}
\begin{split}
\label{AT1+split}
    \mathcal{E}_l(\boldsymbol{u},d) = & \int_{\Omega} \frac{1}{2}\Big((1-d)^2\boldsymbol{\sigma}^+(\boldsymbol{u}) + \boldsymbol{\sigma}^-(\boldsymbol{u})\Big):\boldsymbol{e}(\boldsymbol{u}) \, \text{d}\boldsymbol{x} \\ & + \frac{3G_c}{8}\int_{\Omega}\Big(\frac{d}{l}+l|\nabla d|^2 \Big) \text{d}\boldsymbol{x}.
\end{split}
\end{equation}

\subsubsection{Irreversibility constraint} \label{irreversibility}
Formulation \eqref{AT1+split} is set in a quasi-static regime, with dependency on time appearing only through the boundary conditions. However, the irreversibility condition found in \eqref{VP} implies an history dependency in the minimization problem. Introducing a discretization of the loading process through the pseudo-time steps $\Delta t^n = t^n - t^{n-1}$, and knowing the solution from the previous step, we look for a solution to the minimization problem at every time increment. Using the time discretization, $\dot{d}\geq 0$ can be approximated through a backward differentiation with
\begin{equation}
	d^n \geq d^{n-1}.
\end{equation}
To avoid a cumbersome notation, we denote the solution of the actual increment $(\boldsymbol{u}^n,d^n)$ with $(\boldsymbol{u},d)$.

The variational problem \eqref{VP} now contains 3 constraints, namely $d \geq 0$, $d \leq 1$ and $d \geq d^{n-1}$. With the degradation function chosen as quadratic, $d \leq 1$ is naturally respected~\hl{\mbox{\cite{miehe_phase_2010,kuhn_degradation_2015}}}. However, $d \geq 0$ and $d \geq d^{n-1}$ need to be imposed. Multiple options for enforcing the $d \geq d^{n-1}$ condition are available in the literature, such as \hl{the primal-dual active set method~\mbox{\cite{heister_primal-dual_2015}}}, the penalty method~\hl{\mbox{\cite{gerasimov_penalization_2019,miehe_thermodynamically_2010}}}, \hl{the Lagrange multiplier method~\mbox{\cite{mang_phase-field_2020}}}, the augmented Lagrangian method~\cite{wheeler_augmented-lagrangian_2014}, or the interior-point method~\cite{wambacq_interior-point_2021}. Here we adopt the simple quadratic penalty, as proposed by~\cite{gerasimov_penalization_2019}
\begin{equation}
\label{penalty}
	 P = \frac{\gamma}{2} \int_{\Omega} \langle d - d^{n-1} \rangle_{-}^{2}\: \text{d}\boldsymbol{x},
\end{equation}
where $\gamma$ is a penalty parameter and $d^{n-1}$ is the solution of the phase-field from the previous time increment. By taking $d^0 \geq 0$, the constraint $d \geq 0$ is simultaneously enforced. Adding the penalty term \eqref{penalty} to the energy functional \eqref{AT1+split} yields an unconstrained minimization problem which reads

\begin{equation}
\label{VPu}
    \boldsymbol{u},d = \textrm{arg}\,\textrm{min } \mathcal{E}_l(\boldsymbol{u},d) 
\end{equation}
with the energy functional 
\begin{equation}
\begin{split}
\label{AT1+split+P}
    \mathcal{E}_l(\boldsymbol{u},d) = & \int_{\Omega} \frac{1}{2}\Big((1-d)^2\boldsymbol{\sigma}^+(\boldsymbol{u}) + \boldsymbol{\sigma}^-(\boldsymbol{u})\Big):\boldsymbol{e}(\boldsymbol{u}) \, \text{d}\boldsymbol{x} \\ & + \frac{3G_c}{8}\int_{\Omega}\Big(\frac{d}{l}+l|\nabla d|^2 \Big) \text{d}\boldsymbol{x} + \frac{\gamma}{2} \int_{\Omega} \langle d - d^{n-1} \rangle_{-}^{2}\: \text{d}\boldsymbol{x}.
\end{split}
\end{equation}

This new problem is an approximation of \eqref{VP}, where the minimizers of \eqref{AT1+split+P} converge to the minimizers of \eqref{AT1+split} with $\gamma \rightarrow \infty$. Instead of increasing $\gamma$ until the Karush-Kuhn-Tucker (KKT) conditions are met, as one would typically do with a penalty method, we use the fixed optimal penalty parameter presented in~\cite{gerasimov_penalization_2019}

\begin{equation}
		\gamma = \frac{G_c}{l}\frac{27}{64 \mathtt{TOL^2_{Ir}}},
\end{equation}
where $\mathtt{TOL_{Ir}}$ is a chosen irreversibility tolerance. \hl{In this work, we use $\mathtt{TOL_{Ir}} = 0.01$ as proposed in~\mbox{\cite{gerasimov_penalization_2019}}}. Derived from the notion of optimal phase-field profile, this constant penalty coefficient was shown to lead to a good approximation of the original constrained problem~\cite{gerasimov_penalization_2019}.

\subsubsection{Euler-Lagrange equations and finite element discretization} \label{section:ELd}
We define the function spaces \hl{$\boldsymbol{\mathcal{V}} \coloneqq \{\boldsymbol{v} \in (H^1(\Omega))^\delta : \boldsymbol{v} = \boldsymbol{0} \text{ on }  \Gamma_D \}$} and $\mathcal{W} \coloneqq \hll{\{ w \in H^1(\Omega) \}}$, with $H^1(\Omega)$ the Sobolev space. Applying directional derivatives towards $(\boldsymbol{u},d)$ to \eqref{AT1+split+P}, the Euler-Lagrange equations of problem \eqref{VPu} can be written in their weak form as~\cite{heister_primal-dual_2015,gerasimov_penalization_2019}: %Find $\boldsymbol{u} \in \boldsymbol{\mathcal{V}}$ and $d \in \mathcal{W}$ such that
\begin{subequations}
\label{EL}
\begin{align}
		& \int_{\Omega} \Big( (1-d)^2\boldsymbol{\sigma}^+(\boldsymbol{u}) + \boldsymbol{\sigma}^-(\boldsymbol{u}) \Big):\boldsymbol{e}(\boldsymbol{v}) \, \text{d}\boldsymbol{x} = 0, \quad \forall \boldsymbol{v} \in \boldsymbol{\mathcal{V}}, \label{EL1}\\
		& \int_{\Omega} -w(1-d) \boldsymbol{\sigma}^+(\boldsymbol{u}):\boldsymbol{e}(\boldsymbol{u}) \, \text{d}\boldsymbol{x} + \frac{3G_c}{8}\int_{\Omega} \Big( \frac{w}{l}+2l\nabla d \cdot \nabla w \Big) \text{d}\boldsymbol{x} \nonumber \\ & \quad + \gamma \int_{\Omega} w\langle d - d^{n-1} \rangle_{-}\: \text{d}\boldsymbol{x} = 0, \quad \forall w \in \mathcal{W}. \label{EL2}
\end{align}
\end{subequations}

The equation system \eqref{EL} is usually discretized using a Galerkin finite element method \cite{heister_primal-dual_2015,borden_phase-field_2012,mang_phase-field_2020}. To this end, assuming a 2D polygonal domain, we partition $\Omega$ in a non-degenerate family of quadrilateral elements $\mathcal{Q}^h$, where $h$ is the average element edge size. Using the notation from~\cite{brenner_mathematical_2008}, we introduce the linear finite element space
\begin{equation*}
	\mathcal{V}_h \coloneqq \{ v \in H^1(\Omega): \text{ $v|_Q$ is linear $\forall Q \in \mathcal{Q}^h$}\}.
\end{equation*}
The nodal basis functions of the vector space $\mathcal{V}_h$ are noted
\begin{equation*}
	\{ \varphi_i : 1 \leq i \leq m \},
\end{equation*}
where $m$ is the dimension of the finite element space. To approximate the vector field $\boldsymbol{u}$, and borrowing the deal.II convention~\cite{bangerth_assembling_2002}, we define the $2m$ vector shape functions
\begin{align*}
\boldsymbol{\phi}_0(\boldsymbol{x})
& =
\begin{pmatrix}
\varphi_0(\boldsymbol{x}) \\
0
\end{pmatrix}, \\
\boldsymbol{\phi}_1(\boldsymbol{x})
& =
\begin{pmatrix}
0 \\
\varphi_0(\boldsymbol{x})
\end{pmatrix}, \\
\boldsymbol{\phi}_2(\boldsymbol{x})
& =
\begin{pmatrix}
\varphi_1(\boldsymbol{x}) \\
0
\end{pmatrix}\hll{,} \ldots \\
\boldsymbol{\phi}_{2m}(\boldsymbol{x})
& =
\begin{pmatrix}
0 \\
\varphi_m(\boldsymbol{x})
\end{pmatrix}.
\end{align*}
Collecting the displacement and damage nodal coefficients in respectively $\boldsymbol{U} \in \mathbb{R}^{2m}$ and $\boldsymbol{D} \in \mathbb{R}^m$, the functions $(\boldsymbol{u},d)$ are now approximated with
\begin{equation}
	\boldsymbol{u}_h(\boldsymbol{x}) = \sum_{j = 1}^{2m}\boldsymbol{\phi}_j(\boldsymbol{x})U_j, \quad d_h(\boldsymbol{x}) = \sum_{j = 1}^{m}\varphi_j(\boldsymbol{x}) D_j.
\end{equation}
Using the finite element approximations $\boldsymbol{u}_h \in \mathcal{V}_h \times \mathcal{V}_h$ and $d_h \in \mathcal{V}_h$, the discretized Euler-Lagrange equations can be written as the following residuals
\begin{subequations}
\label{ELd}
\begin{align}
		R_i^u \coloneqq & \int_{\Omega} \Big( (1-d_h)^2\boldsymbol{\sigma}^+(\boldsymbol{u}_h) + \boldsymbol{\sigma}^-(\boldsymbol{u}_h) \Big):\boldsymbol{e}(\boldsymbol{\phi}_i) \, \text{d}\boldsymbol{x} = 0 \label{ELd1}\\
		R_i^d \coloneqq & \int_{\Omega} -(1-d_h)\varphi_i\boldsymbol{\sigma}^+(\boldsymbol{u}_h):\boldsymbol{e}(\boldsymbol{u}_h) \, \text{d}\boldsymbol{x} + \frac{3G_c}{8}\int_{\Omega} \Big( \frac{\varphi_i}{l}+2l\nabla d_h \cdot \nabla \varphi_i \Big) \text{d}\boldsymbol{x} \nonumber \\ & \: + \gamma \int_{\Omega} \varphi_i\langle d_h - d_h^{n-1} \rangle_{-}\: \text{d}\boldsymbol{x} = 0. \label{ELd2}
\end{align}
\end{subequations}

For all Newton-based solvers, the assembly of the Jacobian, or at least part of it, is necessary. The Jacobian associated to \eqref{ELd} can be written in its block structure as 
\begin{subequations}
\label{Jacobian}
\begin{align}
		J^{uu}_{ij} = \frac{\partial R_i^u}{\partial U_j} = & \int_{\Omega} \Big( (1-d_h)^2\boldsymbol{\sigma}^+(\boldsymbol{\phi}_j) + \boldsymbol{\sigma}^-(\boldsymbol{\phi}_j) \Big) :\boldsymbol{e}(\boldsymbol{\phi}_i) \, \text{d}\boldsymbol{x} \label{J1}\\
		J^{ud}_{ij} = \frac{\partial R_i^u}{\partial D_j} = & \int_{\Omega} -2(1-d_h)\varphi_j\boldsymbol{\sigma}^+(\boldsymbol{u}_h):\boldsymbol{e}(\boldsymbol{\phi}_i) \, \text{d}\boldsymbol{x} \label{J2}\\
		J^{du}_{ij} = \frac{\partial R_i^d}{\partial U_j} = & \int_{\Omega} -2(1-d_h)\varphi_i\boldsymbol{\sigma}^+(\boldsymbol{u}_h):\boldsymbol{e}(\boldsymbol{\phi}_j) \, \text{d}\boldsymbol{x} \label{J3}\\
		J^{dd}_{ij} = \frac{\partial R_i^d}{\partial D_j} = &  \int_{\Omega} \varphi_i\varphi_j\boldsymbol{\sigma}^+(\boldsymbol{u}_h):\boldsymbol{e}(\boldsymbol{u}_h) \text{d}\boldsymbol{x} + \frac{3G_c}{8}\int_{\Omega}2l\nabla \varphi_i \cdot \nabla \varphi_j \text{d}\boldsymbol{x} \nonumber \\ &\; + \gamma \int_{\Omega} \varphi_i\varphi_jH^-(d_h - d_h^{n-1}) \: \text{d}\boldsymbol{x}, \label{J4}
\end{align}
\end{subequations}
with the general structure
\begin{equation}
\boldsymbol{J} = 
	\begin{pmatrix}
		\boldsymbol{J}^{uu} & \boldsymbol{J}^{ud} \\
		\boldsymbol{J}^{du} & \boldsymbol{J}^{dd}
	\end{pmatrix}.
\end{equation}

By $\boldsymbol{\sigma}^+(\boldsymbol{\phi}_j)$, we imply the directional derivative of $\boldsymbol{\sigma}^+(\boldsymbol{u}_h)$ towards $U_j$ as thoroughly detailed by~\cite{heister_primal-dual_2015}. Stemming from the derivative of the ramp function $\langle z \rangle_{-}$, $H^-(z)$ is the Heaviside function worth 1 if $z \leq 0$ and worth 0 otherwise. Finally, we denote the discretized Dirichlet conditions $\boldsymbol{\bar{U}}$.

\subsection{Solution algorithms} \label{Solution algorithms}
We present two algorithms commonly used to obtain a solution in phase-field fracture problems: the alternating minimization algorithm and the quasi-monolithic scheme. With these two methods, the Euler-Lagrange equations \eqref{EL} are treated as a set of non-linear equations and only a stationary point of \eqref{AT1+split+P} is obtained. The nature of the stationary point as either a local minimum, saddle point, or local maximum is unknown.

\subsubsection{Alternating minimization algorithm}
As observed in~\cite{bourdin_numerical_2000}, the functional \eqref{AT1} becomes convex in one field when the other is fixed. A stationary point can therefore be easily obtained with Newton's method if we solve towards one field at a time. Based on this fact, the alternating minimization algorithm, also often referred to as the staggered solver, consists in splitting the problem in a kinematic and a damage sub-problem and, as its name implies, alternatively solving them until convergence in both $\boldsymbol{u}$ and $d$. Since the solution is independent from the previous time steps, except of course for the penalty term, the alternating minimization algorithm is an implicit solver.

Applied to the energy functional \eqref{AT1+split+P}, the staggered solver sequentially solves \eqref{ELd1} and \eqref{ELd2} using their respective Jacobian $J^{uu}$ \eqref{J1} and $J^{dd}$ \eqref{J4} through Newton's method. In this work, we adopt the alternating minimization algorithm as presented in~\cite{marigo_overview_2016,gerasimov_penalization_2019}, and described in Algorithm \ref{Stagg}.

\begin{algorithm}
\caption{Alternating minimization scheme}
\label{Stagg}
\hspace*{\algorithmicindent} \textbf{Input:} $\boldsymbol{U}^{n-1}$, $\boldsymbol{D}^{n-1}$, $\boldsymbol{\bar{U}}^n$ on $\Gamma_D$. \\
\hspace*{\algorithmicindent} \textbf{Output:} $\boldsymbol{U}^{n}$, $\boldsymbol{D}^{n}$.
\begin{algorithmic}[1]
\State Initialize $\boldsymbol{U}_0 \coloneqq \boldsymbol{U}^{n-1}$, $\boldsymbol{D}_0 \coloneqq \boldsymbol{D}^{n-1}$, $k = 0$.
\While{$||\boldsymbol{R}^d(\boldsymbol{U}_k,\boldsymbol{D}_k)||_{L^\infty} > \mathtt{TOL}$}
\State Find solution $\boldsymbol{D}_{k+1}$: $\hll{||\boldsymbol{R}^d(\boldsymbol{U}_k,\boldsymbol{D}_{k+1})||_{L^\infty}} \leq \mathtt{TOL}_{IN}$.
\State Find solution $\boldsymbol{U}_{k+1}$: $\hll{||\boldsymbol{R}^u(\boldsymbol{U}_{k+1},\boldsymbol{D}_{k+1})||_{L^\infty}} \leq \mathtt{TOL}_{IN}$.
\State $k \gets k+1$.
\EndWhile
\State $\boldsymbol{U}^{n} \gets \boldsymbol{U}_{k}$, $\boldsymbol{D}^{n} \gets \boldsymbol{D}_{k}$.
\end{algorithmic}
\end{algorithm}

Different stopping criteria are used in the literature, see~\cite{wu_bfgs_2020} for a short review. Here, we use a residual-based convergence \hl{criterion, as shown in Algorithm \mbox{\ref{Stagg}}}. Since the two sub-problems are solved sequentially, it is necessary for the inner Newton methods to be solved to a smaller tolerance to reach the global convergence \hl{criterion}, implying the necessity of selecting $\texttt{TOL}_{IN} < \texttt{TOL}$~\cite{gerasimov_penalization_2019}.

\subsubsection{Quasi-monolithic scheme} \label{Quasi-monolithic scheme}
Based on the identification of the term $(1-d)^2\boldsymbol{\sigma}^+(\boldsymbol{u}):\boldsymbol{e}(\boldsymbol{u})$ as the one rendering the functional \eqref{AT1+split+P} non-convex, \hl{the idea of the quasi-monolithic scheme is to linearize the kinematic sub-problem using a fixed} $\tilde{d}$ instead of the unknown field $d$~\cite{heister_primal-dual_2015}. $\tilde{d}$ is taken as an extrapolation constructed from the solution of previous time steps, with $\tilde{d} \coloneqq \tilde{d}(d^{n-1},d^{n-2})$. \hl{To retain the coupling between the two fields, $\tilde{d}$ is inserted in lieu of $d$ only in equation \mbox{\eqref{EL1}}, rendering the kinematic sub-problem dependent of $\boldsymbol{u}$ only. The damage sub-problem is kept identical to \mbox{\eqref{EL2}}}. However, due to the extrapolation $\tilde{d}$, the quasi-monolithic scheme can be interpreted as a semi-implicit method, therefore explaining the numerical viscosity observed in~\cite{wick_error-oriented_2017}.

Different types of extrapolations are used for $\tilde{d}$, such as the linear extrapolation appearing in the implementation available with~\cite{heister_primal-dual_2015} or the constant extrapolation described in~\cite{mang_phase-field_2020}. In fact, many extrapolation schemes could be developed and tested. However, formulation \eqref{AT1+split+P} has no dependency in time, implying that no guarantee can be made on the quality of a time-based extrapolation of $d$. By replacing the unknown phase-field with a heuristic prediction as done here, one could reasonably expect the performance of this method to be problem-dependent.

To perform a fair but straightforward comparison between the different methods, we adopt the linear extrapolation, as used in~\cite{heister_primal-dual_2015,lee_phase-field_2018,wick_error-oriented_2017} and described in~\hl{\mbox{\cite{wick_multiphysics_2020}}} as
\begin{equation}
	\tilde{d} = d^{n-1}\frac{t^n - t^{n-2}}{t^{n-1} - t^{n-2}} - d^{n-2}\frac{t^n - t^{n-1}}{t^{n-1} - t^{n-2}}.
\end{equation}

With the extrapolated scheme, the residual reads
\begin{equation}
\label{ResQM}
	\begin{split}
		R_i^u \coloneqq & \int_{\Omega} \Big( (1-\tilde{d}_h)^2\boldsymbol{\sigma}^+(\boldsymbol{u}_h) + \boldsymbol{\sigma}^-(\boldsymbol{u}_h) \Big) :\boldsymbol{e}(\boldsymbol{\phi}_i) \, \text{d}\boldsymbol{x} = 0 \\
		R_i^d \coloneqq & \int_{\Omega} -(1-d_h)\varphi_i\boldsymbol{\sigma}^+(\boldsymbol{u}_h):\boldsymbol{e}(\boldsymbol{u}_h) \, \text{d}\boldsymbol{x} \\ & \: + \frac{3G_c}{8}\int_{\Omega} \Big( \frac{\varphi_i}{l}+2l\nabla d_h \cdot \nabla \varphi_i \Big) \text{d}\boldsymbol{x} + \gamma \int_{\Omega} \varphi_i\langle d_h - d_h^{n-1} \rangle_{-}\: \text{d}\boldsymbol{x} = 0 
	\end{split}
\end{equation}
with the Jacobian
\begin{equation}
\label{JacQM}
	\begin{split}
		J^{uu}_{ij} = \frac{\partial R_i^u}{\partial U_j} = & \int_{\Omega} \Big( (1-\tilde{d}_h)^2\boldsymbol{\sigma}^+(\boldsymbol{\phi}_j) + \boldsymbol{\sigma}^-(\boldsymbol{\phi}_j) \Big) :\boldsymbol{e}(\boldsymbol{\phi}_i) \, \text{d}\boldsymbol{x} \\
		J^{ud}_{ij} = \frac{\partial R_i^u}{\partial D_j} = & \;0 \\
		J^{du}_{ij} = \frac{\partial R_i^d}{\partial U_j} = & \int_{\Omega} -2(1-d_h)\varphi_i\boldsymbol{\sigma}^+(\boldsymbol{u}_h):\boldsymbol{e}(\boldsymbol{\phi}_j) \, \text{d}\boldsymbol{x} \\
		J^{dd}_{ij} = \frac{\partial R_i^d}{\partial D_j} = & \int_{\Omega} \varphi_i\varphi_j\boldsymbol{\sigma}^+(\boldsymbol{u}_h):\boldsymbol{e}(\boldsymbol{u}_h) \text{d}\boldsymbol{x} + \frac{3G_c}{8}\int_{\Omega} 2l\nabla \varphi_i \cdot \nabla \varphi_j \text{d}\boldsymbol{x} \\ & + \gamma \int_{\Omega} \varphi_i\varphi_jH^-(d_h - d_h^{n-1}) \: \text{d}\boldsymbol{x}
	\end{split}.
\end{equation}

For every time step $t^n$, the solution $\boldsymbol{U} \coloneqq \{\boldsymbol{u}^n,d^n\}$ is obtained with the Newton method using the modified residual $\boldsymbol{R}$ \eqref{ResQM} and Jacobian $\boldsymbol{J}$ \eqref{JacQM}. In this case, the Newton method would consist in solving the linear system
\begin{equation}
	\boldsymbol{J}_k \boldsymbol{\delta U}_k = -\boldsymbol{R}_k
\end{equation}
and updating the solution $\boldsymbol{U}_k$ through
\begin{equation}
	\boldsymbol{U}_{k+1} = \boldsymbol{U}_k + \boldsymbol{\delta U}_k
\end{equation}
until the convergence \hl{criterion} is reached. Here, we use the residual-based convergence \hl{criterion} $||\boldsymbol{R}(\boldsymbol{U}_k)||_{L^\infty} \leq \mathtt{TOL}$.

\section{New solution algorithms} \label{New solution algorithms}
\subsection{A modified Newton method with inertia correction and energy line-search}
When minimizing a function $\mathcal{E}(\boldsymbol{u})$, a positive definite Hessian, or Jacobian, guarantees the identified Newton direction $\boldsymbol{\delta u}_{k}$ to be one of descent. A descent direction implies the existence of a solution $\boldsymbol{u}_{k+1} \coloneqq \boldsymbol{u}_{k} + \alpha_k \boldsymbol{\delta u}_{k}$, with the step-size parameter $\alpha_k > 0$, such that $\mathcal{E}(\boldsymbol{u}_{k+1}) < \mathcal{E}(\boldsymbol{u}_{k})$. Therefore, if the Jacobian is positive definite and a line-search procedure is used to identify a satisfying $\alpha_k$, the convergence to a local minimum is theoretically guaranteed~\cite{nocedal_numerical_2006}.

To preserve this convergence property when solving phase-field models, we propose to use a modified Newton method that ensures the positive definiteness of the Jacobian through inertia corrections, inspired by~\cite{wachter_implementation_2006}. The inertia of an $m \times m$ matrix is expressed with the triplet $(\chi^+,\chi^-,\chi^0)$, where $\chi^+$, $\chi^-$ and $\chi^0$ are respectively the number of positive, negative and null eigenvalues, counting the multiplicities. Consequently, the inertia of a positive definite Jacobian of size $m \times m$ should be $(m,0,0)$. Since the inertia is available as an output with symmetric indefinite linear solvers such as the ones found in the Intel® Math Kernel Library (MKL) and Harwell Subroutine Library (HSL), it suffices to verify that $\chi^+ = m$ before taking a step in the search direction. If the Jacobian is not positive definite, we correct it by adding multiples of the identity matrix, yielding the modified Newton step~\cite{nocedal_numerical_2006}
\begin{equation}
\label{modifiedNewton}
	(\boldsymbol{J}_k + \tau_k \boldsymbol{I}) \boldsymbol{\delta U}_k = -\boldsymbol{R}_k,
\end{equation}
where $\tau_k > 0$ is a scalar parameter and $\boldsymbol{I}$ is the identity matrix. To choose $\tau_k$, we adopt the inertia correction algorithm used in the optimization software IPOPT~\cite{wachter_implementation_2006}. The modified Newton method with inertia correction is presented in Algorithm~\ref{MN}. As described in Algorithm~\ref{MN}, we first attempt to obtain a search direction without correction. \hl{If an incorrect inertia is detected, a first small correction $\bar{\tau}>0$ is tested. If the correction is still insufficient, the correction is multiplied by $\kappa^+ > 1$ until all eigenvalues are positive. If a correction is needed in a subsequent iteration, the last used value of $\tau$ is reused as a starting point, with a slight decrease by $\kappa^- \in [0,1]$. $\bar{\tau}^{min} > 0$ and $\bar{\kappa}^+ > 1$ denote the smallest correction accepted and the increase used for the first iteration with a non-zero correction. For example, the default parameters used in IPOPT are $\kappa^+ = 8$, $\kappa^- = \frac{1}{3}$, $\bar{\kappa}^+ = 100$, $\bar{\tau} = 10^{-4}$, and $\bar{\tau}^{min} = 10^{-20}$~\mbox{\cite{wachter_implementation_2006}}.}

\hl{As seen in \mbox{Algorithm~\ref{MN}}, the $\kappa^+$, $\kappa^-$, $\bar{\kappa}^+$, $\bar{\tau}$ and $\bar{\tau}^{min}$ parameters control the amplitude of the Jacobian regularization and the rapidity with which the amplitude is increased. Hence, they should be chosen to allow \mbox{Algorithm~\ref{MN}} to identify the $\tau_k$ rendering the modified Hessian positive definite through a minimum number of factorization, while still trying to minimize the amplitude of the modification made to the original Hessian. Since the inertia correction is a heuristic method, the $\kappa^+$, $\kappa^-$, $\bar{\kappa}^+$, $\bar{\tau}$ and $\bar{\tau}^{min}$ are indeed numerical parameters that could be tuned. However, as shown in  \mbox{Section~\ref{Numerical results}}, significant and consistent gains in performance are obtained with this method without any tuning and by simply taking the default parameters of IPOPT.}

Therefore, for every time step $t^n$, the solution $\boldsymbol{U} \coloneqq \{\boldsymbol{u}^n, d^n\}$ is obtained by identifying a search direction $\boldsymbol{\delta U}_k$ using Algorithm \ref{MN}, with residual $\boldsymbol{R}_k$ \eqref{ELd} and the full Jacobian \eqref{Jacobian}, and updating the solution $\boldsymbol{U}_k$ with
\begin{equation}
	\boldsymbol{U}_{k+1} = \boldsymbol{U}_k + \alpha_k \boldsymbol{\delta U}_k
\end{equation}
until the convergence \hl{criterion} is reached. Once again, we use the residual-based convergence \hl{criterion} $||\boldsymbol{R}(\boldsymbol{U}_k)||_{L^\infty} \leq \mathtt{TOL}$. $\alpha_k \in [0,1]$ is identified through a standard Armijo backtracking line-search~\hl{\mbox{\cite{nocedal_numerical_2006}} applied to the energy functional~\mbox{\eqref{AT1+split+P}}, as described in Algorithm~\mbox{\ref{Armijo}} where $\rho \in [0,1]$ is the contraction factor}.

The described modified Newton method presents many theoretical and numerical advantages over the solvers found in the phase-field literature. First, since the Jacobian is modified only when necessary, as opposed to the staggered and quasi-monolithic solvers, the modified Newton method preserves the theoretical quadratic convergence rate of the Newton method in as many time steps as possible. Once the Jacobian is modified, the convergence rate remains linear~\cite{nocedal_numerical_2006}. Note also that the corrections appear naturally within the method and, \hl{contrarily to the solvers proposed in~\mbox{\cite{farrell_linear_2017,storvik_accelerated_2021}}, no switch between algorithms is required}. Second, as opposed to methods relying only on the residual and Jacobian, the use of a line-search on the energy functional ensures that the solution is at least a local minimum. Altogether, these multiple advantages make the proposed modified Newton method a potentially highly efficient and robust fully monolithic solver for phase-field models.

\begin{algorithm}[]
\caption{Modified Newton method with inertia correction}
\label{MN}
\hspace*{\algorithmicindent} \textbf{Input:} $\boldsymbol{J}_k$, $\boldsymbol{R}_k$, $\tau_{k-1}$, $\kappa^+$, $\kappa^-$, $\bar{\kappa}^+$, $\bar{\tau}$, $\bar{\tau}^{min}$. \\
\hspace*{\algorithmicindent} \textbf{Output:} $\boldsymbol{\delta U}_k$, $\tau_{k}$.
\begin{algorithmic}[1]
\State Initialize $\tau^0 = 0$, $i = 0$.
\State Factorize $(\boldsymbol{J}_k + \tau^0 \boldsymbol{I})$ and obtain the inertia $(\chi^+,\chi^-,\chi^0)$.
\If{the modified Jacobian is not positive definite, i.e. $\chi^+ \neq m$}
	
	\If{$\tau_{k-1}$ = 0}
		\State Set $\tau^1 = \bar{\tau}$
	\Else
		\State Set $\tau^1 = \text{max}(\bar{\tau}^{min},\kappa^- \tau_{k-1})$
	\EndIf
	\State Factorize $(\boldsymbol{J}_k + \tau^1 \boldsymbol{I})$ and obtain the inertia $(\chi^+,\chi^-,\chi^0)$.
	\State $i \gets i+1$
	\While{the modified Jacobian is not positive definite, i.e. $\chi^+ \neq m$}
		\If{$\tau_{k-1}$ = 0}
			\State Set $\tau^{i+1} = \bar{\kappa}^+ \tau^i$
		\Else
			\State Set $\tau^{i+1} = \kappa^+ \tau^i$
		\EndIf
	\State Factorize $(\boldsymbol{J}_k + \tau^{i+1} \boldsymbol{I})$ and obtain the inertia $(\chi^+,\chi^-,\chi^0)$.
	\State $i \gets i+1$
	\EndWhile
\EndIf
\State Solve $(\boldsymbol{J}_k + \tau^i \boldsymbol{I}) \boldsymbol{\delta U} = -\boldsymbol{R}_k$ for $\boldsymbol{\delta U}$.
\State $\boldsymbol{\delta U}_k \gets \boldsymbol{\delta U}$, $\tau_{k} \gets \tau^i$.
\end{algorithmic}
\end{algorithm}

\begin{algorithm}
\caption{\hl{Armijo backtracking line-search}}
\label{Armijo}
\hspace*{\algorithmicindent} \textbf{Input:} $\boldsymbol{U}_k$, $\boldsymbol{\delta U}_k$, $\rho$. \\
\hspace*{\algorithmicindent} \textbf{Output:} $\alpha_k$.
\begin{algorithmic}[1]
\State Initialize $\alpha^0 = 1$, $i = 0$.
\While{$\mathcal{E}_l(\boldsymbol{U}_k + \alpha^i \boldsymbol{\delta U}_k) > \mathcal{E}_l(\boldsymbol{U}_k)$}
\State Set $\alpha^{i+1} = \rho \alpha^i$.
\State $i \gets i+1$.
\EndWhile
\State $\alpha_k \gets \alpha^i$.
\end{algorithmic}
\end{algorithm}

\subsection{An improved quasi-monolithic scheme using an extrapolation correction loop}
The quasi-monolithic scheme as described in Section~\ref{Quasi-monolithic scheme} is a semi-implicit method, implying that, depending on the step size, the solution can be delayed or completely inaccurate (see~\cite{wick_error-oriented_2017} for an example). The lag in the solution can be explained by the extrapolation $\tilde{d}$ underestimating the rapidly evolving damage field when too large time steps are used. Due to the absence of time continuity of the damage field, large time steps can cause the extrapolation to quickly diverge from the actual solution.

The simplest strategy to lessen the lag induced by the extrapolation is to use uniformly refined time steps. On one hand, this method usually yields a more accurate solution, since, with the step size converging to zero, the time-based extrapolation can reasonably be expected to converge to the implicit solution. On the other hand, the uniformly refined time steps imply additional and unnecessary calculations when no crack is propagating. One could consider the use of a time adaptive scheme, but once again, the lack of continuity in time of the damage field hinders the potential robustness of such methods. To the best of our knowledge, no such time adaptive scheme is available for the quasi-monolithic scheme in the literature.

Another strategy to reduce the lag in the evolution of the damage field, as proposed in~\cite{wick_multiphysics_2020}, is to use an additional loop in which the extrapolation is updated. The idea is that, by updating the extrapolation inside a single time step, $\tilde{d}$ should get closer and eventually converge to the actual damage solution $d$.

Rather than using a constant and maybe insufficient number of updates, we propose to continue the iterating process until the damage field reaches a convergence \hl{criterion}. The solution is defined as converged when the updating of the extrapolation no longer influences the phase-field solution, within a certain tolerance, as
\begin{equation}
\label{time_conv}
	||d_{k} - d_{k-1}||_{L^2} \leq \mathtt{TOL_{QM}}.
\end{equation}

\hl{The final quasi-monolithic method is detailed in Algorithm~\mbox{\ref{Quasi-mono}}. As described, for every time steps, a first extrapolation $\widetilde{\boldsymbol{D}}_{1}$ is constructed from the phase-field solution of the two previous time steps, $\boldsymbol{D}^{n-1}$ and $\boldsymbol{D}^{n-2}$. Using $\widetilde{\boldsymbol{D}}_{1}$, a first solution $\{\boldsymbol{U}_1, \boldsymbol{D}_1\}$ is computed solving the modified residual~\mbox{\eqref{ResQM}}. The extrapolation $\widetilde{\boldsymbol{D}}_{k+1}$ is then updated using the two most recent phase-field iterates, $\boldsymbol{D}_{k}$ and $\boldsymbol{D}_{k-1}$, and the solution $\{\boldsymbol{U}_k, \boldsymbol{D}_k\}$ re-calculated until the stopping criterion~\mbox{\eqref{time_conv}} is respected. Note that no additional precaution was taken to enforce $\boldsymbol{D}^n \leq 1$.}

\hl{The proposed extrapolation correction loop is heuristic since $d$ has no regularity in time, but applying Algorithm~\mbox{\ref{Quasi-mono}} every time steps allows the semi-implicit quasi-monolithic scheme to converge to the implicit solution without a uniform time step refinement or pre-defined number of extrapolation correction, as shown in~\mbox{\ref{Appendix}}.}

\begin{algorithm}
\caption{Quasi-monolithic scheme with extrapolation correction loop}
\label{Quasi-mono}
\hspace*{\algorithmicindent} \textbf{Input:} $\boldsymbol{U}^{n-1}$, $\boldsymbol{D}^{n-1}$, $\boldsymbol{D}^{n-2}$, $\boldsymbol{\bar{U}}^n$ on $\Gamma_D$. \\
\hspace*{\algorithmicindent} \textbf{Output:} $\boldsymbol{U}^{n}$, $\boldsymbol{D}^{n}$.
\begin{algorithmic}[1]
\State Initialize $\boldsymbol{U}_0 \coloneqq \boldsymbol{U}^{n-1}$, $\boldsymbol{D}_{0} \coloneqq \boldsymbol{D}^{n-1}$, $\boldsymbol{D}_{-1} \coloneqq \boldsymbol{D}^{n-2}$, $k = 1$.
\State Compute extrapolation $\widetilde{\boldsymbol{D}}_{1} \coloneqq f(\boldsymbol{D}_{0}, \boldsymbol{D}_{-1})$.
\State Find solution $\{\boldsymbol{U}_{1},\boldsymbol{D}_{1}\}$: $\boldsymbol{R}(\boldsymbol{U}_1,\boldsymbol{D}_{1}, \widetilde{\boldsymbol{D}}_{1}) \leq \mathtt{TOL}$.
\While{$||\boldsymbol{D}_{k} - \boldsymbol{D}_{k-1}||_{L^2} > \mathtt{TOL_{QM}}$}
\State Compute extrapolation $\widetilde{\boldsymbol{D}}_{k+1} \coloneqq f(\boldsymbol{D}_{k}, \boldsymbol{D}_{k-1})$.
\State Find solution $\{\boldsymbol{U}_{k+1},\boldsymbol{D}_{k+1}\}$: $\hll{||\boldsymbol{R}(\boldsymbol{U}_{k+1},\boldsymbol{D}_{k+1},\boldsymbol{\widetilde{D}}_{k+1})||_{L^\infty}} \leq \mathtt{TOL}$.
\State $k \gets k+1$.
\EndWhile
\State $\boldsymbol{U}^{n} \gets \boldsymbol{U}_{k}$, $\boldsymbol{D}^{n} \gets \boldsymbol{D}_{k}$.
\end{algorithmic}
\end{algorithm}

\section{Numerical results} \label{Numerical results}

In this section, we present benchmarks representative of numerical experiments found in the phase-field literature. For all tests and for each algorithm, we compare the resultant crack paths, the force-displacement responses of the structure, the number of iterations required, and the computation time. We define an iteration as every time a correction is made to the solution $(\boldsymbol{u}^n,d^n)$. This definition implies that iterations do not have the same associated computational cost depending on the algorithm. Nevertheless, we consider the total number of iterations to inform on the performance of the algorithms when combined with the computation time.

Only 2D geometries are considered and \hl{plane} strain is assumed. For every geometry, a local mesh refinement with a maximum size of $h = \frac{l}{5}$ is used in the expected fracture zone to avoid a numerical overestimation of the toughness~\cite{tanne_crack_2018,bourdin_variational_2008,mesgarnejad_validation_2015}. Pre-existing cracks are modeled as notches in the geometry and mesh. All meshes were generated using the commercial finite element software ABAQUS. The boundary conditions are strongly enforced by isolating the degrees of freedom subject to the Dirichlet condition and reducing the linear system. \hl{All computations are displacement-driven with constant displacement increments.} The integrals are approximated using the Gauss quadrature with four integration points. 

For all three methods, the convergence tolerance on the residual is set to $\mathtt{TOL} = 10^{-4}$, \hl{as in~\mbox{\cite{gerasimov_penalization_2019,gerasimov_line_2016,gerasimov_stochastic_2020}}}. The tolerance of the inner Newton methods of the alternating algorithm is set to $\mathtt{TOL}_{IN} = 10^{-5}$. In order to propose the fairest comparison with the quasi-monolithic scheme, the tolerance $\mathtt{TOL}_{QM}$ is adjusted for each experiment to ensure convergence to the implicit solution while minimizing the computation time. For the inertia correction scheme of the modified Newton solver, we take the default IPOPT parameters with $\kappa^+ = 8$, $\kappa^- = \frac{1}{3}$, $\bar{\kappa}^+ = 100$, $\bar{\tau} = 10^{-4}$ and $\bar{\tau}^{min} = 10^{-20}$ \cite{wachter_implementation_2006}. \hl{For the Armijo backtracking line-search used with the modified Newton solver, we take the contraction factor $\rho = \frac{1}{2}$. We recall that, as suggested in~\mbox{\cite{gerasimov_penalization_2019}}, an irreversibility tolerance of $\mathtt{TOL_{Ir}} = 0.01$ is taken for the penalty coefficient.}

The finite element method and the numerical algorithms presented are implemented using the Julia language~\cite{bezanson_julia_2017}. For all three methods, linear systems are solved with the Intel® MKL Pardiso direct sparse solver. All computations are performed on a workstation with an Intel® Core(TM) i7-6700K CPU @ 4.00GHz and 32 GB RAM.

\subsection{Single-edge notched plate under tensile loading} \label{SENP-tensile}

\begin{figure}[]
	\begin{subfigure}{.5\textwidth}
		\centering
		\includegraphics[width=\textwidth]{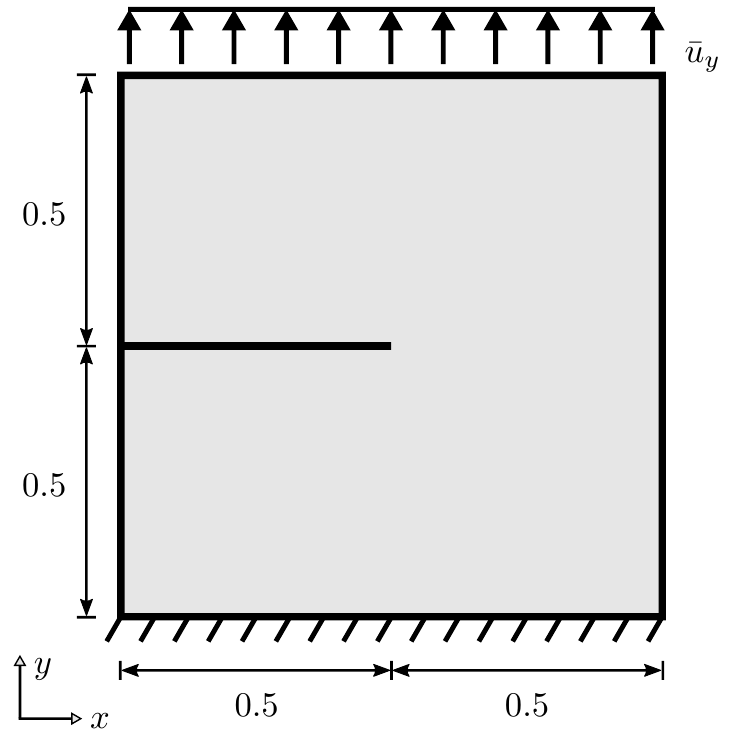}
		\caption{}
	\end{subfigure}
	\begin{subfigure}{.5\textwidth}
		\centering
		\includegraphics[width=\textwidth]{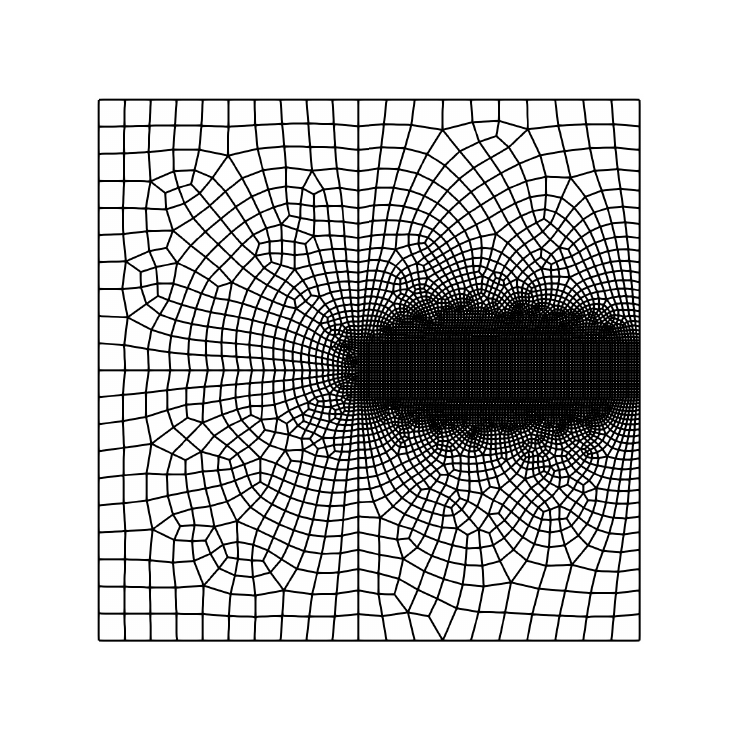}
		\caption{}
	\end{subfigure}
	\caption{Single edge notched plate under tension (SENP-tension) test. (a) Schematic of the geometry and boundary conditions. The lower edge of the plate is fixed (i.e. $\bar{u}_x = \bar{u}_y = 0$) while only a vertical displacement of $\bar{u}_y = 0.01$ mm is prescribed on the upper edge. The plate has unit thickness.  All units are in mm. (b) Finite element discretization of the geometry.}
	\label{fig:Tensile}
\end{figure}

As a first test case, we consider the simple notched square plate under uniaxial tension (SENP-tensile). Widely used as a reference test in the literature~\cite{martinez-paneda_phase_2018,heister_primal-dual_2015,miehe_phase_2010,ambati_review_2015}, the geometry and loading generate unstable crack propagation. With an implicit scheme, the crack is expected to fully extend across the plate within a single time increment. Therefore, this test allows to assess the capacity of the algorithms to capture fast propagating cracks.

The geometry, the boundary conditions and the mesh used for this test are presented in Figure~\ref{fig:Tensile}. The material properties selected are $E = 210$ GPa, $\nu = 0.3$, $G_c = 2.7$ N/mm and $l = 0.024$ mm. A total vertical displacement of 0.01 mm is applied in 50 time steps, implying uniform increments of $\Delta \bar{u} = 2\times 10^{-4}$ mm. The finite element mesh contains 6524 nodes and 6431 quadrilateral elements. A tolerance of $\mathtt{TOL}_{QM} = 0.01$ was found sufficient for the extrapolated scheme to converge to the implicit solution.

\begin{figure}[!tb]
	\centering
	\includegraphics[width=\textwidth]{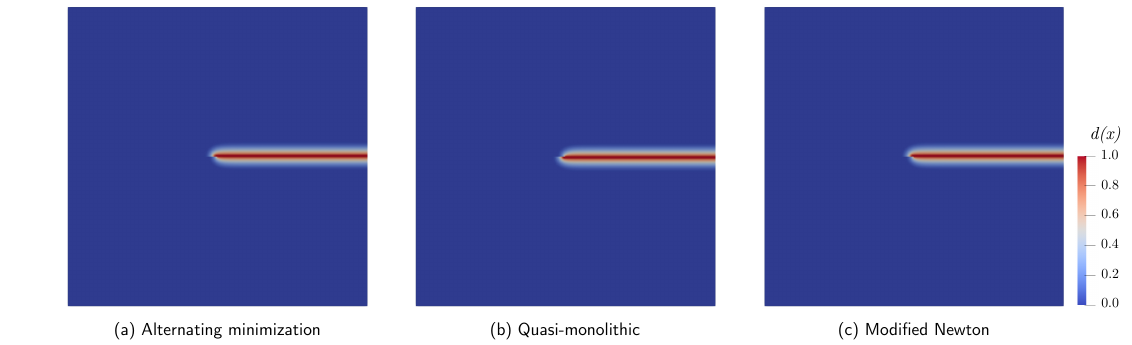}
\caption{Comparison of the final crack patterns ($\bar{u}_y = 0.01$ mm) for the three studied algorithms on the undeformed geometry of the SENP-tension test. Crack paths obtained with the three methods are in agreement. The crack produced by the quasi-monolithic (b) scheme has initiated from below the notch, as opposed to the two other methods where the crack initiated above the notch. The difference remains however negligible.} 
\label{fig:CrackTensile}
\end{figure}

\begin{figure}[p!]
	\centering
	\includegraphics[width=0.65\textwidth]{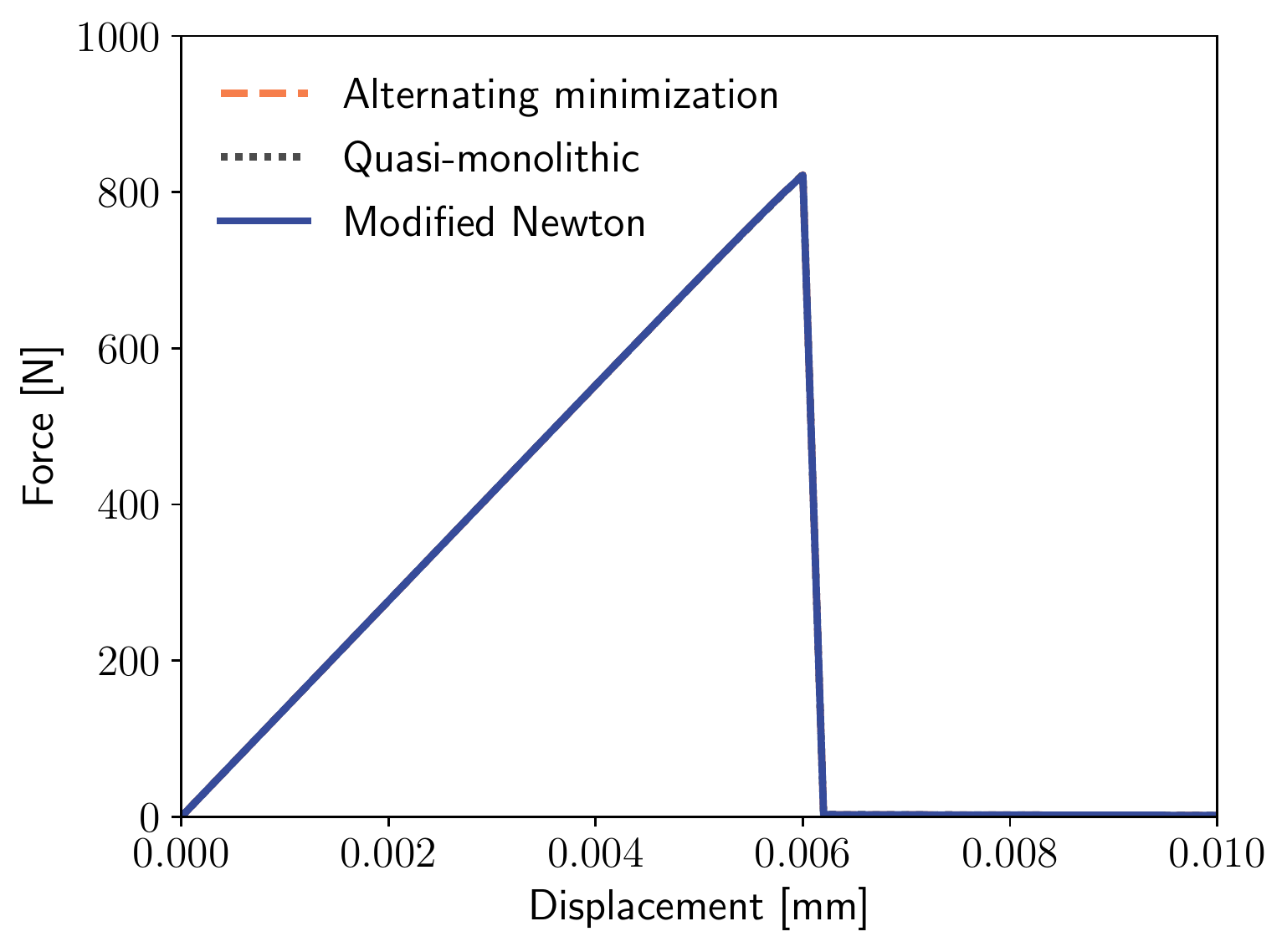}
	\caption{Load-displacement curves obtained with the different solvers on the SENP-tension test. The three algorithms produce identical solutions.}
	\label{fig:TensileFD}
\end{figure}

Figure~\ref{fig:CrackTensile} shows the crack paths obtained with \hl{the alternating minimization, quasi-monolithic, and modified Newton algorithms} for the SENP-tensile tests. As depicted, all three algorithms converge to the expected solution, which in this case is a straight crack from the pre-crack to the right edge. Figure~\ref{fig:TensileFD} presents the corresponding force-displacement responses obtained with the solvers. The three methods yielded identical force-displacement curves, with a linear response until the peak load, followed by a sudden drop to zero at $\bar{u}_y = 6.2\times 10^{-3}$ mm. The results obtained with the three solvers confirm that they are able to capture the instantaneous propagation and that the quasi-monolithic scheme coupled with the extrapolation correction loop, as presented in Algorithm \ref{Quasi-mono}, converges to the implicit solution.

\begin{figure}[p!]
	\centering
	\includegraphics[width=0.65\textwidth]{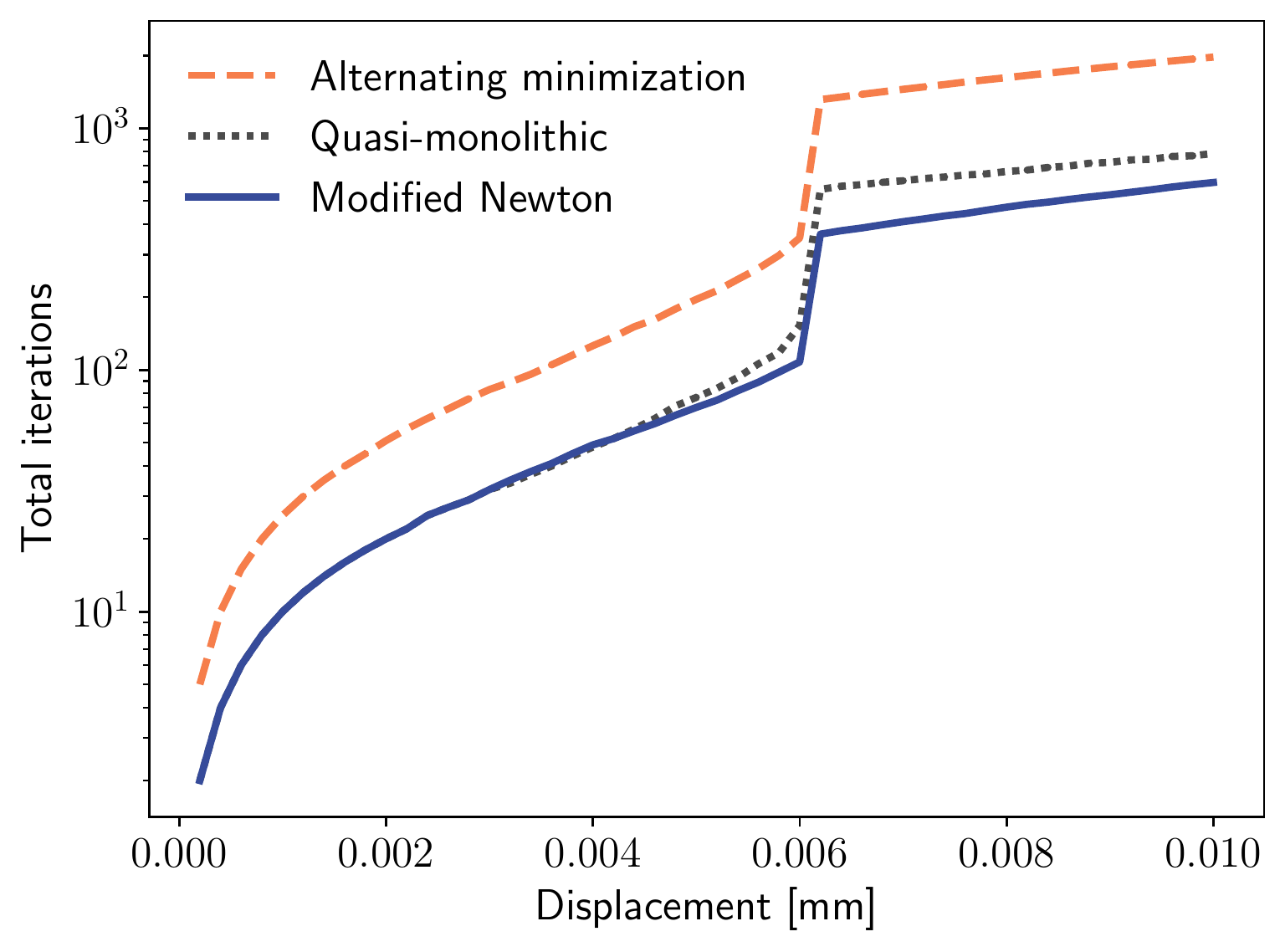}
	\caption{Total number of iterations required for the different algorithms to converge on the SENP-tension test. The modified Newton solver appears as the most efficient, while the extrapolated scheme also shows a reduction of the number of iterations when compared to the alternating minimization algorithm.}
	\label{fig:TensileIT}
\end{figure}

Figure~\ref{fig:TensileIT} shows the total number of iterations required for each algorithm to converge with the evolution of the boundary conditions. At time step 31 ($\bar{u}_y = 6.2\times 10^{-3}$ mm), where the crack is propagating, the modified Newton solver requires only 257 iterations, as opposed to the 406 and 963 needed by the quasi-monolithic scheme and the alternating minimization algorithm, respectively. Table \ref{table:Tensile} summarizes the number of iterations and the computation time required by the three algorithms for the SENP-tensile test. \hl{As can be seen, the modified Newton method needed to correct the Hessian in only 30 iterations out of the 597 total iterations. Therefore, only 5 seconds was spent computing the inertia corrections. Furthermore, the decrease in the number of iterations obtained with the modified Newton solver yielded a reduction of the computation time by factors of 1.53 and 1.19, when compared to the staggered and extrapolated solvers. However, damage propagation is present in only one increment. Consequently, this benchmark does not fully expose the efficiency of the proposed method.}

\begin{table}[p!]
	\centering
	\caption{Comparison of the iterations and computation time required for the studied algorithms to solve the SENP-tension benchmark. \hl{For the modified Newton method, the number of iterations requiring inertia correction (IC) and the total time spent computing the inertia corrections are indicated.}}
	\label{table:Tensile}
	\addtolength{\leftskip} {-2cm}
	\addtolength{\rightskip}{-2cm}
	\setlength{\tabcolsep}{6pt}
	\renewcommand{\arraystretch}{1.2}
	\begin{tabular}{cccccc}
	\hline
	                   & Max it. / inc.  & Total it.  & IC it.  & Total time [s] & IC time [s]           \\ \hline
	Alternating minimization & 963            & 1972  & -   & 514            & -             \\
	Quasi-monolithic         & 406             & 787   & -   & 400             & -             \\
	Modified Newton          & 257             & 597    & \hl{30}  & 336             & \hl{5}             \\ \hline
	\end{tabular}
\end{table}

\subsection{Single-edge notched plate under shear loading} \label{SENP-shear}

\begin{figure}[!tb]
	\begin{subfigure}{.5\textwidth}
		\centering
		\includegraphics[width=\textwidth]{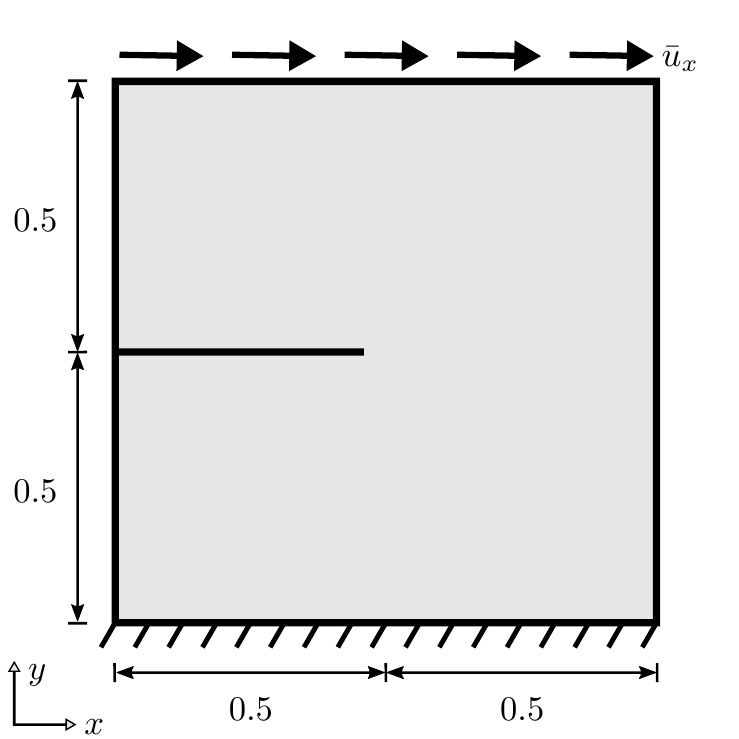}
		\caption{}
	\end{subfigure}
	\begin{subfigure}{.5\textwidth}
		\centering
		\includegraphics[width=\textwidth]{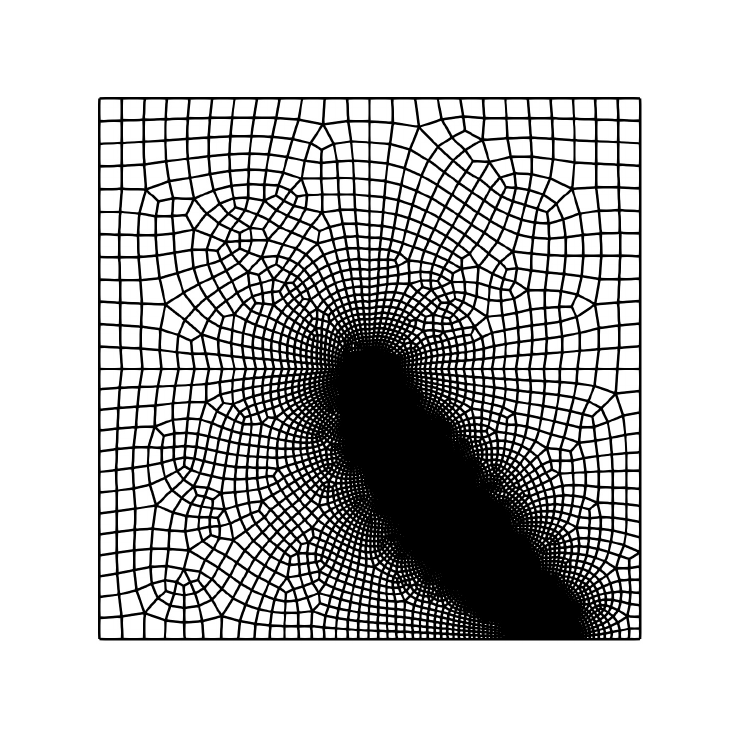}
		\caption{}
	\end{subfigure}
	\caption{Single edge notched plate under shear (SENP-shear) test. (a) Illustration of the geometry and boundary conditions. The lower edge of the plate is fixed (i.e. $\bar{u}_x = \bar{u}_y = 0$). \hl{A horizontal} displacement of $\bar{u}_x = 0.015$ mm is prescribed on the upper edge while maintaining $\bar{u}_y = 0$. The plate has a unit thickness. The units are in mm. (b) Finite element discretization.}
	\label{fig:Shear}
\end{figure}

The notched square plate under shear loading (SENP-shear) is another widely adopted test in the phase-field community since it allows the verification of the energy splitting method~\cite{heister_primal-dual_2015,miehe_phase_2010,gerasimov_penalization_2019,ambati_review_2015,bilgen_crack-driving_2019,dijk_strain_2020}. \hl{This test is also interesting when studying the efficiency of solvers since, once initiated, the crack propagation is stable.} From the latter numerical experiments found in the literature, it appears that only~\cite{gerasimov_penalization_2019} reported results for the $AT_1$ model. We used the same parameters to allow a comparison with the results from~\cite{gerasimov_penalization_2019}, with $E = 210$ GPa, $\nu = 0.3$, $G_c = 2.7$ N/mm and $l = 0.01$ mm. The geometry, the boundary conditions and the mesh are presented in Figure~\ref{fig:Shear}. A total horizontal displacement of 0.015 mm is applied in 50 equal increments. The mesh is made of 23341 nodes assembled in 23205 elements. In this case, a smaller tolerance of $\mathtt{TOL}_{QM} = 5\times 10^{-3}$ was required because a delay of the initiation and an oscillation of the reaction force were observed for a greater tolerance (results not shown).

\begin{figure}[b!]
	\centering
	\includegraphics[width=\textwidth]{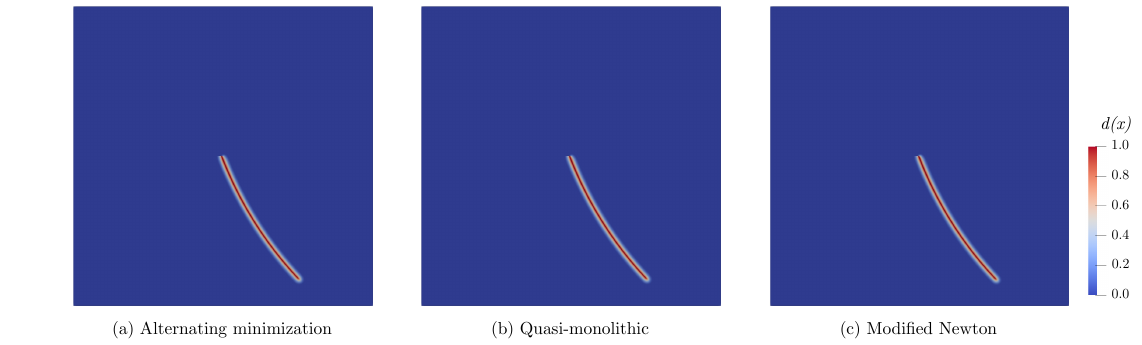}
\caption{Comparison of the final crack paths at $\bar{u}_x = 0.015$ mm for all three studied methods on the undeformed geometry of the SENP-shear test. All three crack patterns are virtually identical.}
\label{fig:CrackShear}
\end{figure}

\begin{figure}[p!]
	\centering
	\includegraphics[width=0.65\textwidth]{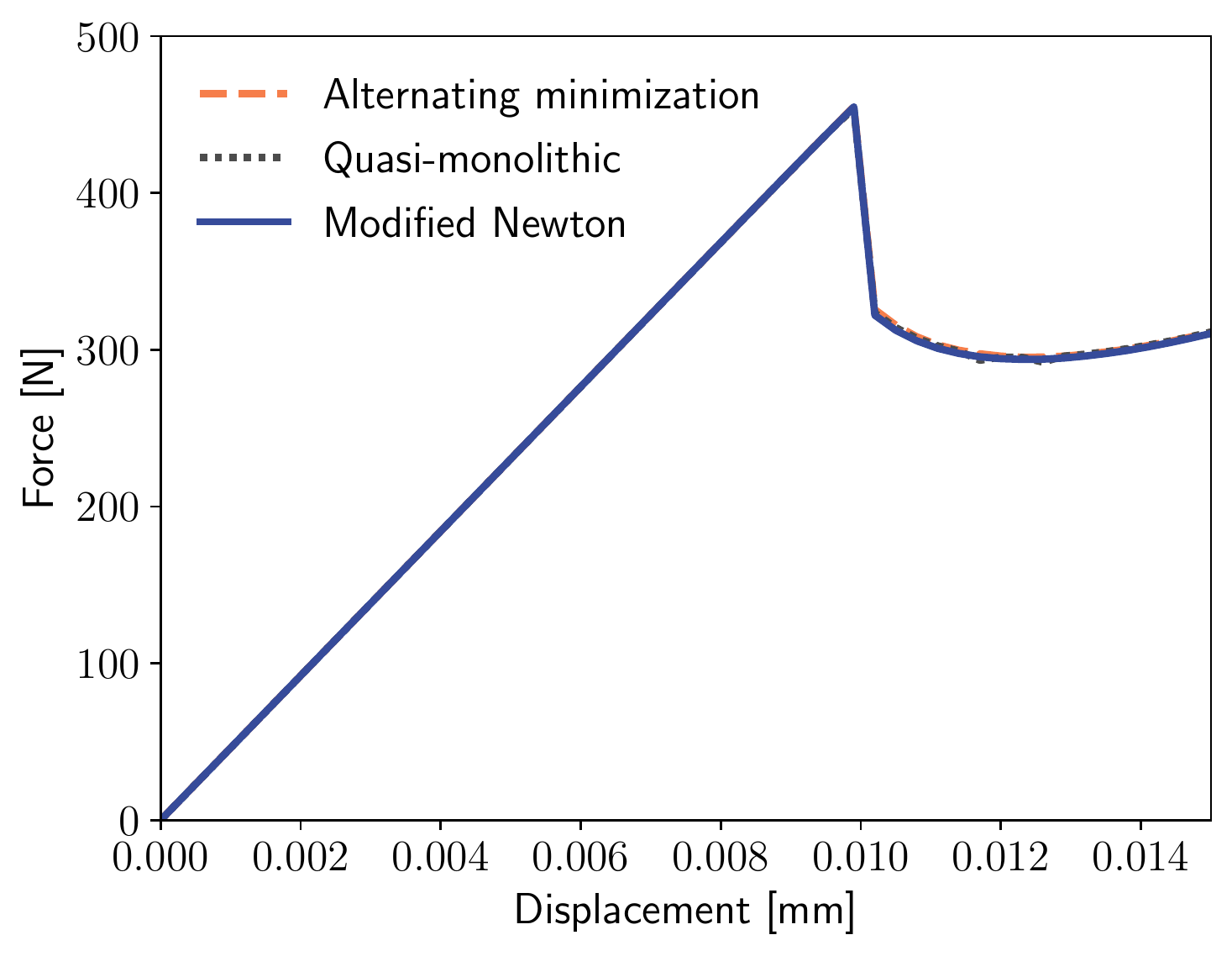}
	\caption{Load versus displacement obtained with the three solvers for the SENP-shear test. The three solutions coincide.}
	\label{fig:ShearFD}
\end{figure}

Figure~\ref{fig:CrackShear} shows the crack paths obtained for the SENP-shear test. As expected and consistently with the results found in the literature (see for example~\cite{ambati_review_2015}), the cracks initiate from the pre-crack and converge in a curved trajectory towards the lower right corner. The cracks yielded by the three solvers are virtually identical. Figure~\ref{fig:ShearFD} presents the load-displacement curves obtained with the three solvers for the SENP-shear test. The responses of the three solvers coincide and match with that of~\cite{gerasimov_penalization_2019}. The difference in peak force observed between the results of Figure~\ref{fig:ShearFD} and the results of~\cite{gerasimov_penalization_2019} could be explained by their use of $P_1$ triangular elements and a mesh size of $h = \frac{l}{4}$.

\begin{figure}[p!]
	\centering
	\includegraphics[width=0.65\textwidth]{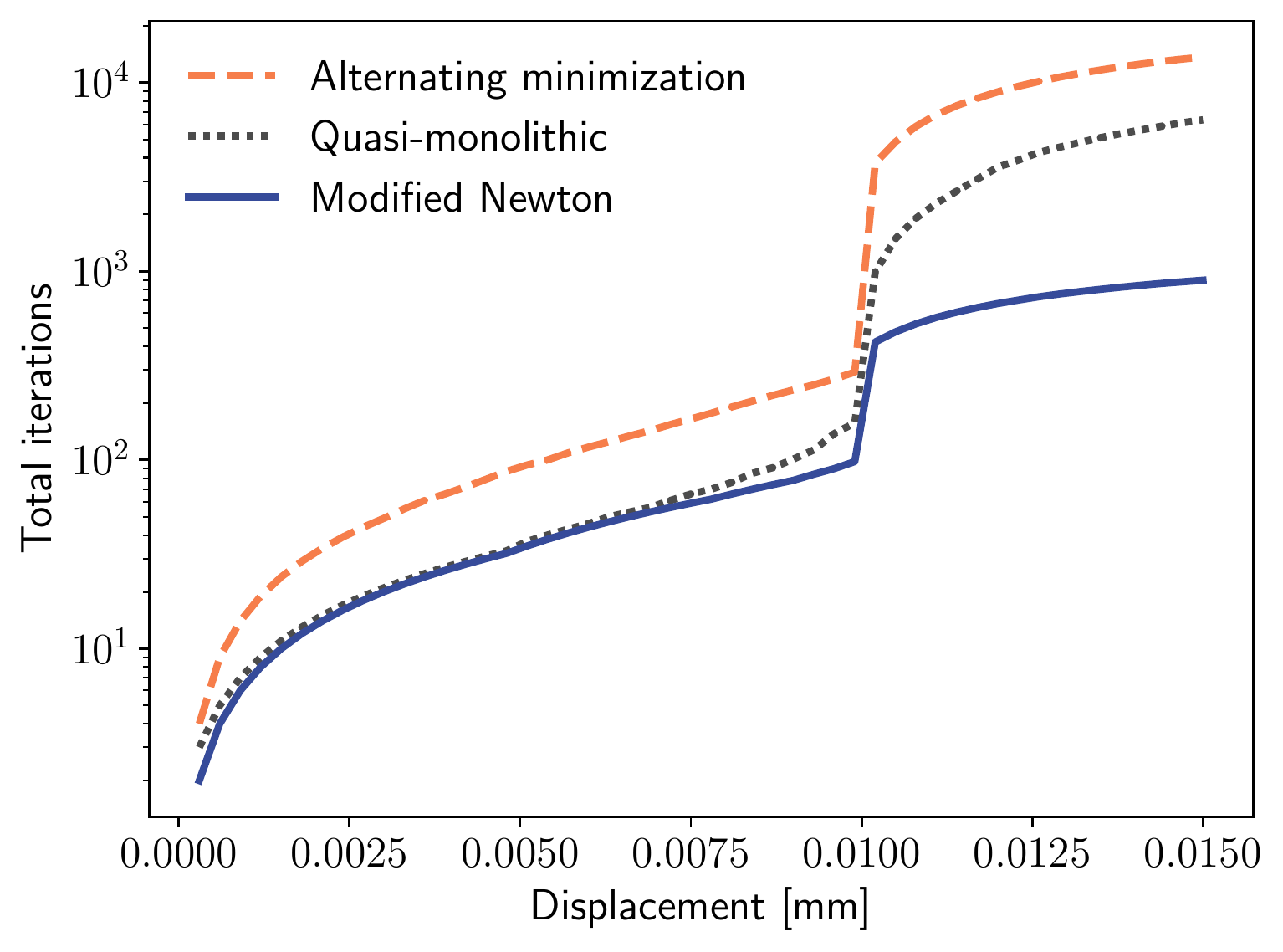}
	\caption{Cumulative number of iterations required by the algorithms depending on the applied displacement in the SENP-shear test. The modified Newton solver clearly necessitates fewer iterations when the damage starts propagating.}
	\label{fig:ShearIT}
\end{figure}

The cumulative number of iterations required by the three solvers is presented in Figure~\ref{fig:ShearIT}. The efficiency of the modified Newton method is clearly illustrated, with a total number of iterations an order of magnitude lower than the quasi-monolithic and alternating schemes. We also note that the iteration totals required by the alternating and modified Newton methods are of the same order of magnitude than that of~\cite{wambacq_interior-point_2021}. Table~\ref{table:Shear} summarizes the computational performances of the three methods for the SENP-shear test. As presented, the total number of iterations required by the modified Newton solver is respectively 15 and 7 times lower when compared to the staggered and the extrapolated scheme. On our workstation, the iteration reduction meant a decrease in computation time by factors of respectively 9.25 and 6.40. \hl{Additionally, with the modified Newton scheme, the Hessian needed to be modified in only 13 iterations, implying that only 9 seconds, out of the total 1557, was spent computing the corrections.} Finally, one can observe from Figure~\ref{fig:ShearIT} that the quasi-monolithic scheme is as efficient as the monolithic modified Newton method prior to the damage propagation. However, the efficiency significantly decreases once the crack starts propagating.

\begin{table}[p!]
\centering
\caption{Comparison of the computational performances of the three algorithms on the SENP-shear test. \hl{For the modified Newton method, the number of iterations requiring inertia correction (IC) and the total time spent computing the inertia corrections are indicated.}}
\label{table:Shear}
\addtolength{\leftskip} {-2cm}
\addtolength{\rightskip}{-2cm}
\setlength{\tabcolsep}{6pt}
\renewcommand{\arraystretch}{1.2}
\begin{tabular}{cccccc}
\hline
Method                   & Max it. / inc.  & Total it.  & IC it.  & Total time [s] & IC time [s]           \\ \hline
Alternating minimization & 3454            & 13654  & -   & 14403            & -             \\
Quasi-monolithic         & 838             & 6349   & -   & 9961             & -             \\
Modified Newton          & 324             & 897    & \hl{13}  & 1557             & \hl{9}             \\ \hline
\end{tabular}
\end{table}

\begin{figure}[p!]
	\centering
	\includegraphics[width=0.65\textwidth]{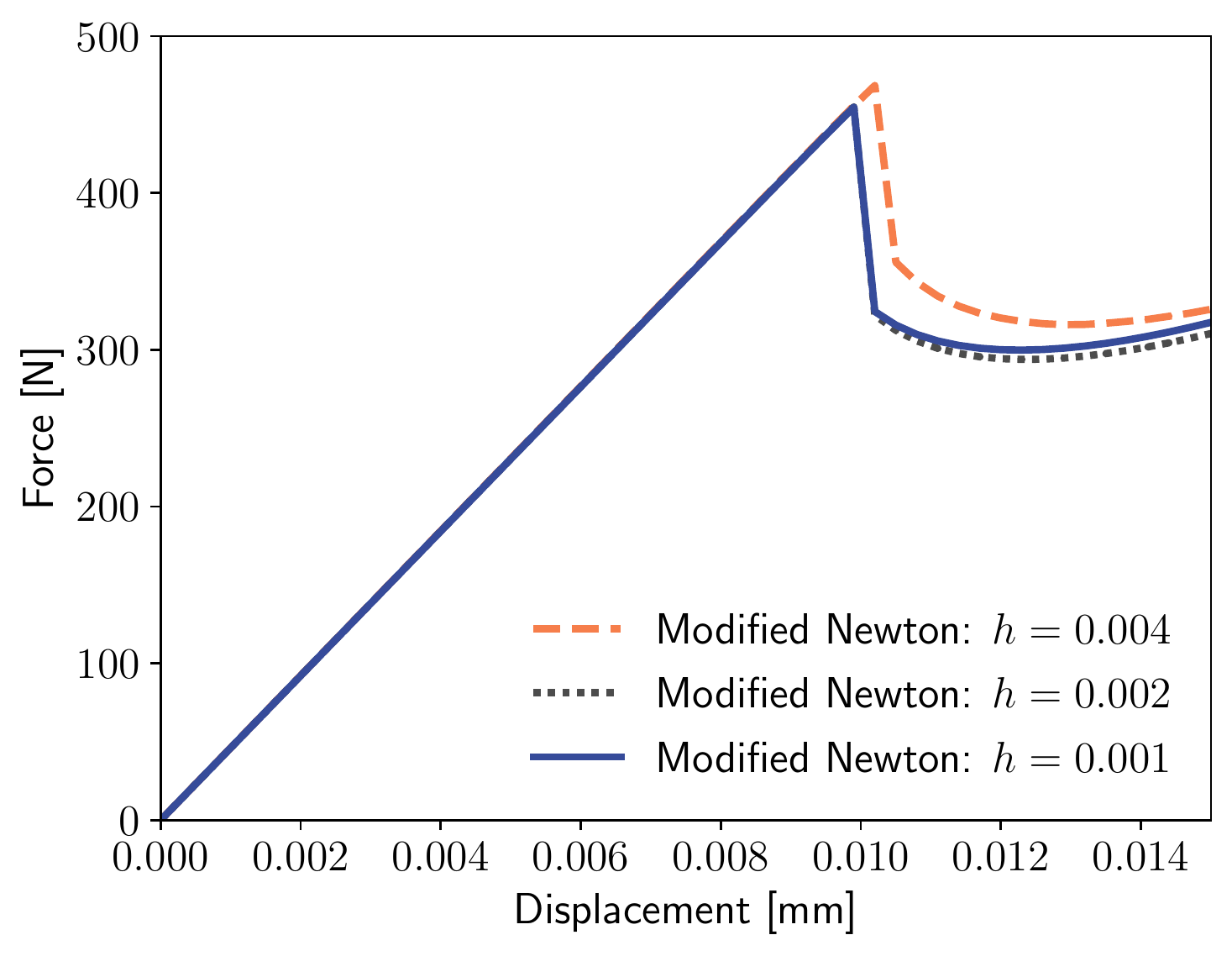}
	\caption{\hl{Load versus displacement obtained with the modified Newton solver on the SENP-shear test for three levels of mesh refinement.}}
	\label{fig:ShearFD_refined}
\end{figure}

\begin{figure}[p!]
	\centering
	\includegraphics[width=0.65\textwidth]{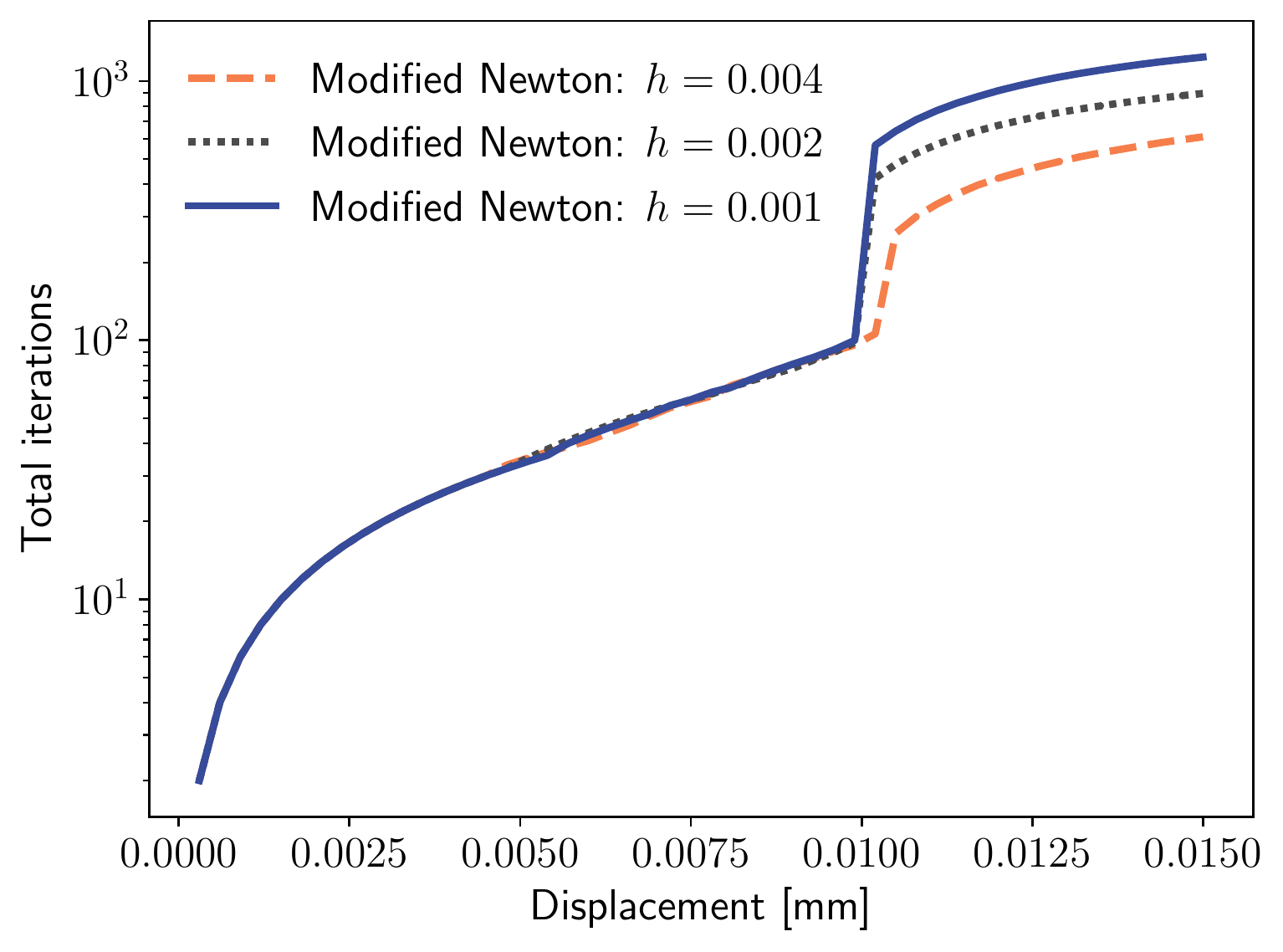}
	\caption{\hl{Cumulative number of iterations required by the modified Newton solver depending on the applied displacement in the SENP-shear test for three levels of mesh refinement.}}
	\label{fig:ShearIT_refined}
\end{figure}

\hl{A mesh refinement study was performed to evaluate the robustness of the modified Newton solver with respect to the spatial discretization. Figure~\mbox{\ref{fig:ShearFD_refined}} compares the force-displacement solutions obtained with the modified Newton algorithm on the SENP-shear test for local mesh refinements in the expected fracture zone of $h = \{ \frac{l}{10}, \frac{l}{5}, \frac{2l}{5} \}$, implying respectively $h = \{ 0.001, 0.002, 0.004 \}$ mm. As seen in Figure~\mbox{\ref{fig:ShearFD_refined}}, the solution is slightly different with $h = 0.004$ mm. This behavior is expected since, as already mentioned, it is well-known that a mesh with $h \leq \frac{l}{5}$ should be used in the fracture zone. Figure~\mbox{\ref{fig:ShearIT_refined}} presents the total number of iterations required for the modified Newton algorithm to converge depending on the mesh refinement. It appears that the number of iterations slightly increases when refining the mesh. The increase in iterations can be partly explained by the initial crack propagation becoming more brutal with finer meshes, thus making the solution more difficult to obtain. Nevertheless, the increase in iterations required by the solver for finer meshes is negligible when compared to the gain in efficiency obtained over the staggered and extrapolated schemes.}

\subsection{Three-point bending test}

\begin{figure}[]
	\begin{subfigure}{.5\textwidth}
		\centering
		\includegraphics[width=\textwidth]{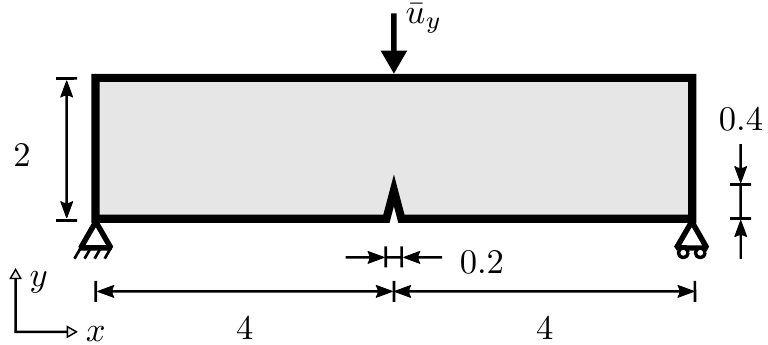}
		\caption{}
	\end{subfigure}
	\begin{subfigure}{.5\textwidth}
		\centering
		\includegraphics[width=\textwidth]{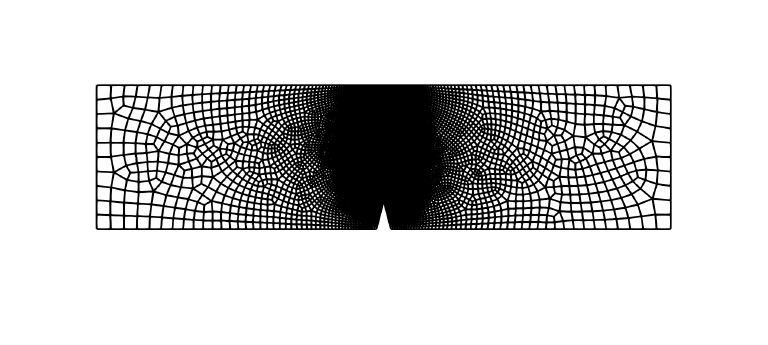}
		\caption{}
	\end{subfigure}
	\caption{Three-point bending of a notched beam test. (a) Schematic of the geometry and boundary conditions. The lower left node is fixed (i.e. $\bar{u}_x = \bar{u}_y = 0$), while the lower right node is only supported vertically and free in $x$ (i.e. $\bar{u}_y = 0$). A vertical displacement of $\bar{u}_y = 0.1$ mm is prescribed on the center node of the upper edge. The beam has a unit thickness. All units are in mm. (b) Finite element discretization of the geometry.}
	\label{fig:3Point}
\end{figure}

Let us now consider the classic three-point bending of a notched beam as presented in~\cite{ambati_review_2015}, \hl{where a mode-I crack first propagates brutally before growing in a stable manner}. The geometry, the loading and the discretization used are depicted in Figure~\ref{fig:3Point}. We adopt the same material properties as in~\cite{ambati_review_2015}, with $E = 20.8$ GPa, $\nu = 0.3$ and $G_c = 0.54$ N/mm. It should be noted, however, that the simulations performed in~\cite{ambati_review_2015} were made using the $AT_2$ model. In order to have equivalent mechanical properties for the $AT_1$ model, the characteristic length is converted using the solution for a homogeneous bar under uniaxial tension~\cite{tanne_crack_2018}. The conversion yields a characteristic length of $l = 0.101$ mm. A total displacement of 0.1 mm discretized in 50 equal increments is applied as illustrated in Figure~\ref{fig:3Point}. 11181 nodes and 11031 elements are used. As in the SENP-tension test, a tolerance of $\mathtt{TOL}_{QM} = 0.01$ was found to be sufficient for the quasi-monolithic scheme to converge to the implicit solution.

\begin{figure}[!tb]
	\centering
	\includegraphics[width=\textwidth]{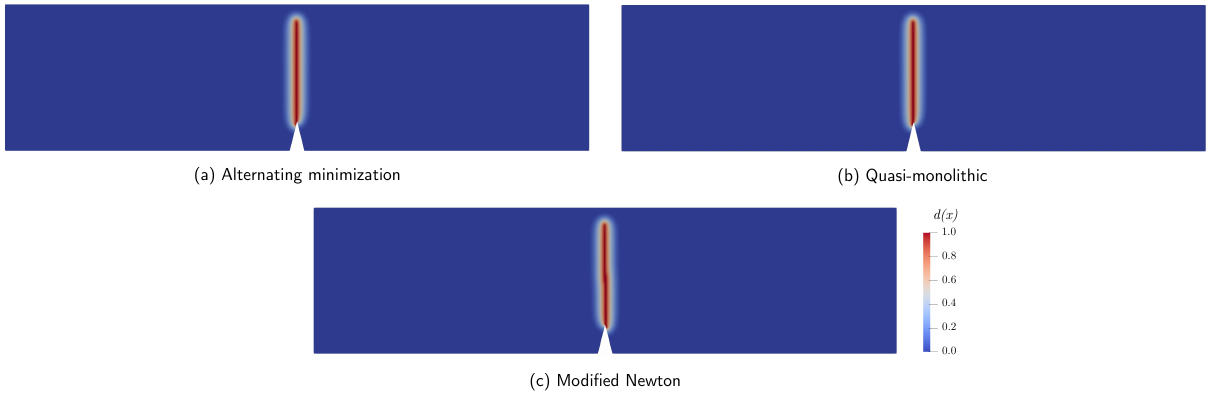}
\caption{Final crack paths for the three methods on the notched beam under three-point bending for a total deflection of $\bar{u}_y = 0.1$ mm. The three cracks are closely similar, however, the crack produced by the modified Newton solver (c) initiated from the right-hand side of the notch, as opposed to the alternating minimization (a) and quasi-monolithic (b) schemes where the crack initiated to the left-hand side.}
\label{fig:Crack3Point}
\end{figure}

\begin{figure}[p!]
\centering
	\centering
	\includegraphics[width=0.65\textwidth]{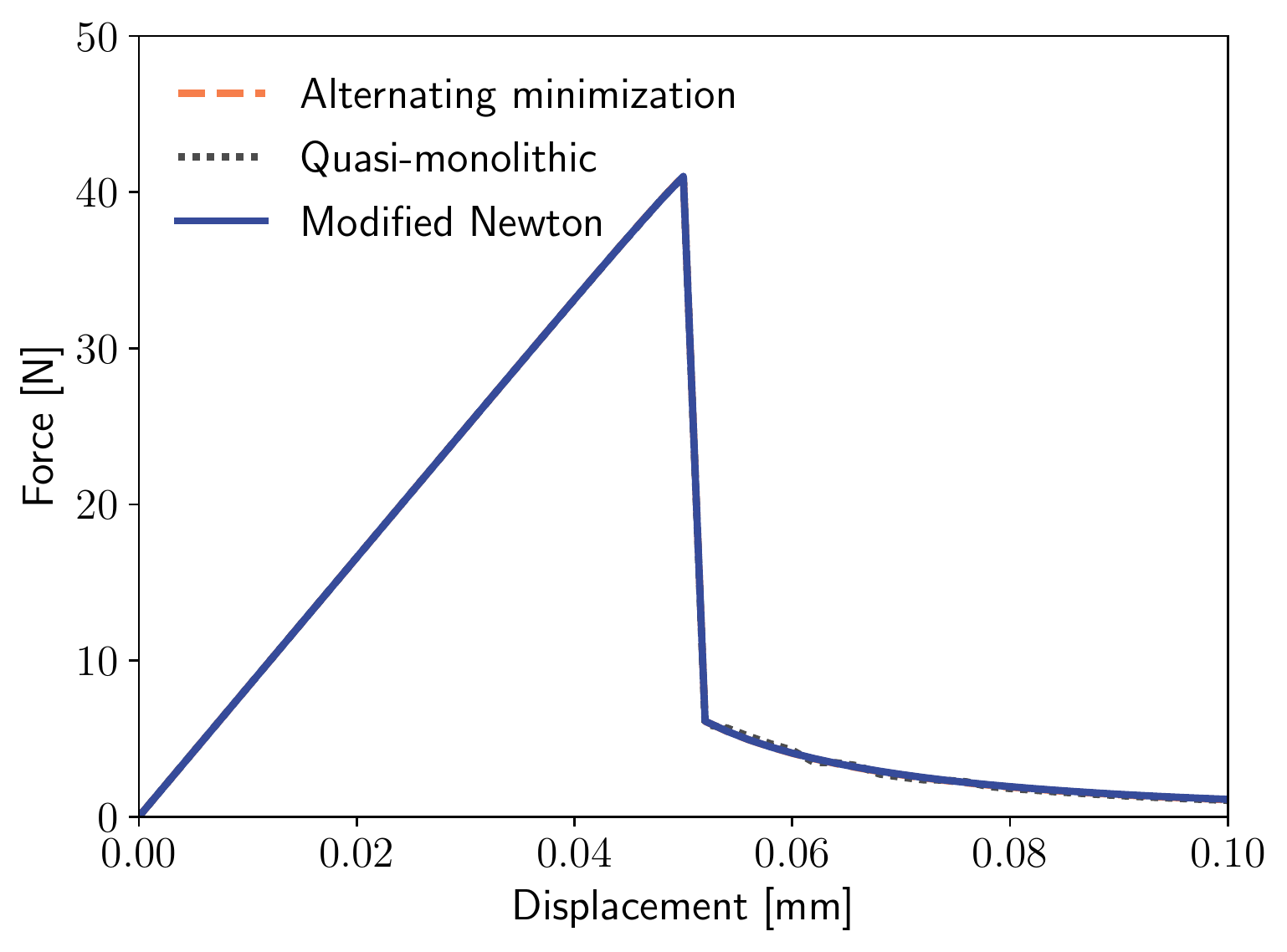}
	\caption{Load-deflection curve of the three-point bending experiment for the different algorithms. Again, all three solutions coincide.}
	\label{fig:3PointFD}
\end{figure}

The resulting crack patterns for the three-point bending test are shown in Figure~\ref{fig:Crack3Point}. As expected for a mode-I fracture, the cracks for all three solvers propagate from the notch to the upper edge in a straight line. Once again, they are virtually identical with only a negligible offset for the solution obtained through the modified Newton solver. Figure~\ref{fig:3PointFD} presents the load-deflection curves obtained for the three-point bending test with the three methods. The solutions are identical and in good agreement with that reported in~\cite{ambati_review_2015}.

\begin{figure}[p!]
	\centering
	\includegraphics[width=0.65\textwidth]{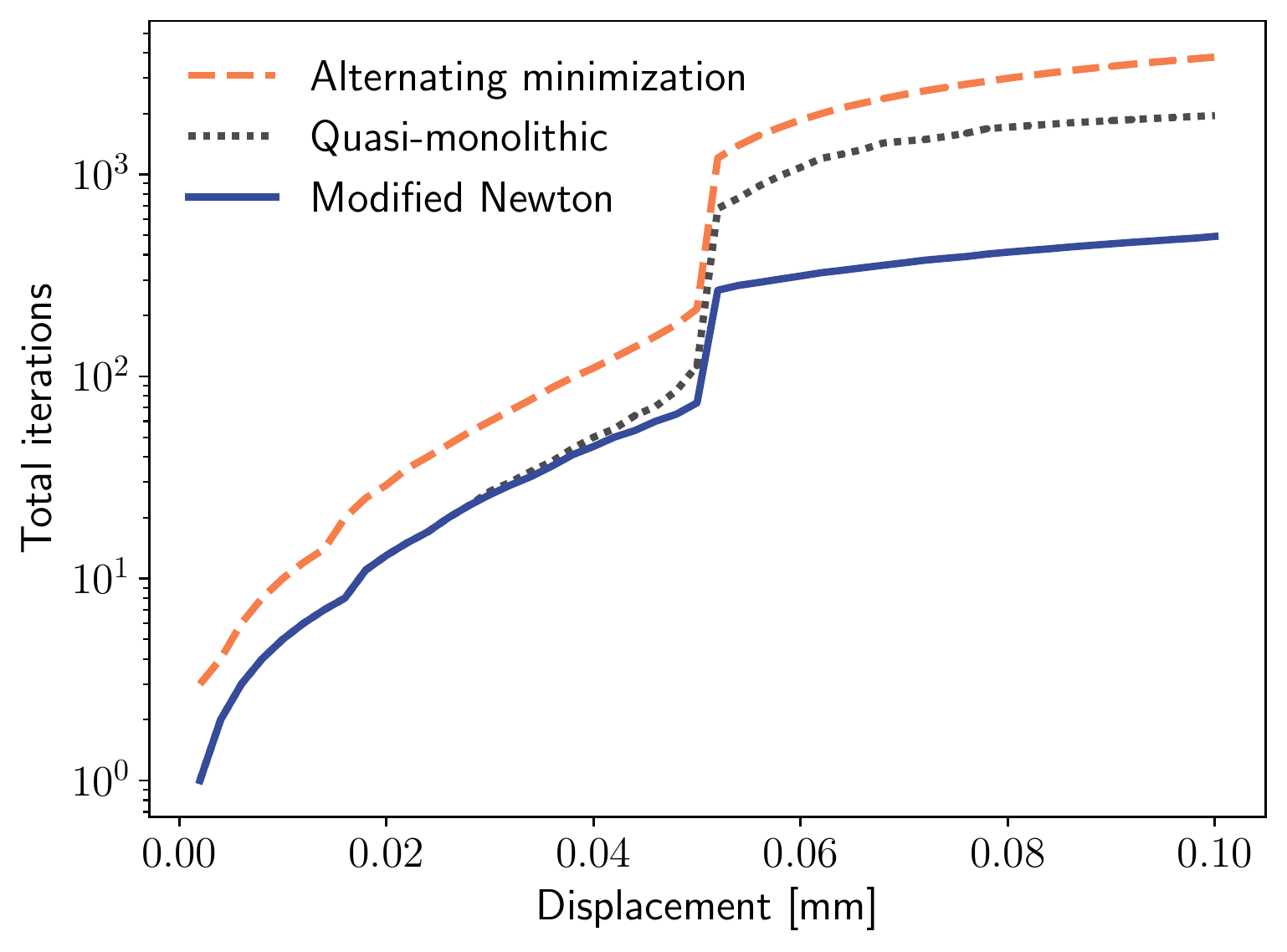}
	\caption{Total number of iterations required by the methods depending on the deflection of the notched beam. Before the crack propagation, the quasi-monolithic scheme is as efficient as the modified Newton method. Once the crack starts propagating, the modified Newton method is clearly the most efficient solver.}
	\label{fig:3PointIT}
\end{figure}

Figure~\ref{fig:3PointIT} presents the cumulative iterations required by the three solvers to obtain a solution to the phase-field problem. As illustrated, the quasi-monolithic and the modified-Newton methods have a similar efficiency before the propagation of damage. However, once the crack starts propagating, the modified Newton algorithm clearly outperforms the alternating and quasi-monolithic schemes. Table \ref{table:3Point} shows the total number of iterations, the maximum number of iterations in an increment and the calculation time required by the three algorithms. The modified Newton solver requires a maximum of only 193 iterations, while the extrapolated and alternating schemes require 565 and 989 iterations, respectively. This is a reduction in the number of iterations by factors of $\approx 3$ and $\approx 5$, which translates to an acceleration of the computation by factors of 3.33 and 4.18, respectively. \hl{Once again, we note that inertia corrections were used in only 70 iterations. The calculations associated to the identification of the correction represented $\approx 4\%$ of the computation time.}

\begin{table}[p!]
\centering
\caption{Computational performances of the different studied methods on the three-point bending test. \hl{For the modified Newton method, the number of iterations requiring inertia correction (IC) and the total time spent computing the inertia corrections are indicated.}}
\label{table:3Point}
\addtolength{\leftskip} {-2cm}
\addtolength{\rightskip}{-2cm}
\setlength{\tabcolsep}{6pt}
\renewcommand{\arraystretch}{1.2}
\begin{tabular}{cccccc}
\hline
Method                   & Max it. / inc.  & Total it.  & IC it.  & Total time [s] & IC time [s]           \\ \hline
Alternating minimization & 989            & 3804  & -   & 1969            & -             \\
Quasi-monolithic         & 565             & 1953   & -   & 1567             & -             \\
Modified Newton          & 193             & 493    & \hl{70}  & 471             & \hl{19}             \\ \hline
\end{tabular}
\end{table}

\subsection{L-shaped panel test}

\begin{figure}[]
	\begin{subfigure}{.5\textwidth}
		\centering
		\includegraphics[width=\textwidth]{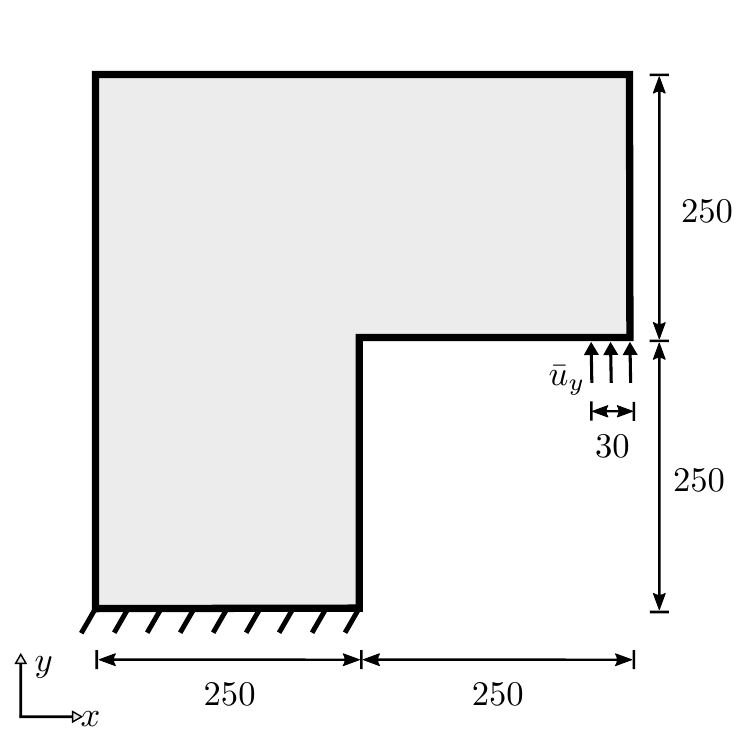}
		\caption{}
	\end{subfigure}
	\begin{subfigure}{.5\textwidth}
		\centering
		\includegraphics[width=\textwidth]{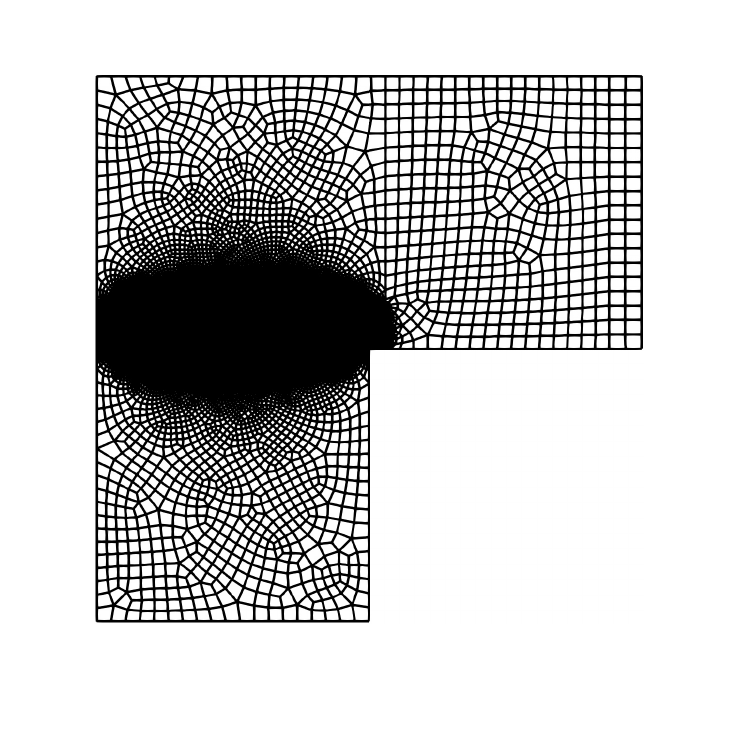}
		\caption{}
	\end{subfigure}
\caption{L-shaped panel test. (a) Representation of the geometry and boundary conditions. The lower edge is fixed (i.e. $\bar{u}_x = \bar{u}_y = 0$) and a vertical displacement of $\bar{u}_y = 1.0$ mm is prescribed as indicated on the figure. The panel has a thickness of 100 mm.  The units are in mm. (b) Finite element discretization of the geometry.}
\label{fig:LShaped}
\end{figure}

The concrete L-shaped panel with mixed mode crack propagation from~\cite{winkler_traglastuntersuchungen_2001} is another popular experiment used to evaluate and validate phase-field models~\cite{wu_bfgs_2020,mang_phase-field_2020,wick_error-oriented_2017,ambati_review_2015,mesgarnejad_validation_2015}. We adopt the geometry, loading and material properties from~\cite{mesgarnejad_validation_2015} for comparison purposes. The geometry, boundary conditions and mesh are presented in Figure~\ref{fig:LShaped}. The material parameters are $E = 21.85$ GPa, $\nu = 0.18$, $G_c = 0.095$ N/mm and $l = 3.125$ mm. A total vertical displacement of 1.0 mm is applied as described in Figure~\ref{fig:LShaped}. 50 equal increments are used to discretize the loading process. The finite element mesh is made of 48984 nodes and 48850 quadrilateral elements. A tolerance of $\mathtt{TOL}_{QM} = 5\times 10^{-3}$ is used for the results presented for the quasi-monolithic scheme, but as will be shown in the following, no convergence to the implicit solution was obtained.

\begin{figure}[!tb]
	\centering
	\includegraphics[width=\textwidth]{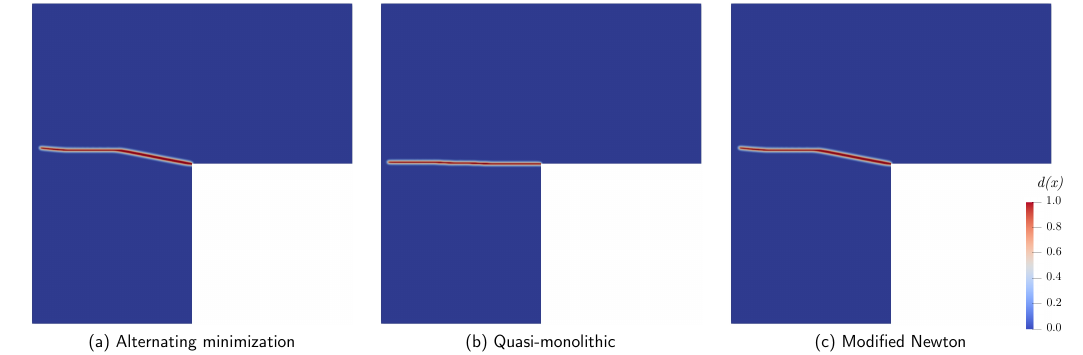}
\caption{Crack paths obtained for the three algorithms after applying a total vertical displacement of $\bar{u}_y = 1.0$ mm. The cracks obtained with the alternating algorithm (a) and the modified Newton method (c) are identical. The solution produced by the extrapolated scheme (b) appears to be incorrect. The undeformed geometry is shown.}
\label{fig:CrackLShaped}
\end{figure}

\begin{figure}[p!]
\centering
	\centering
	\includegraphics[width=0.63\textwidth]{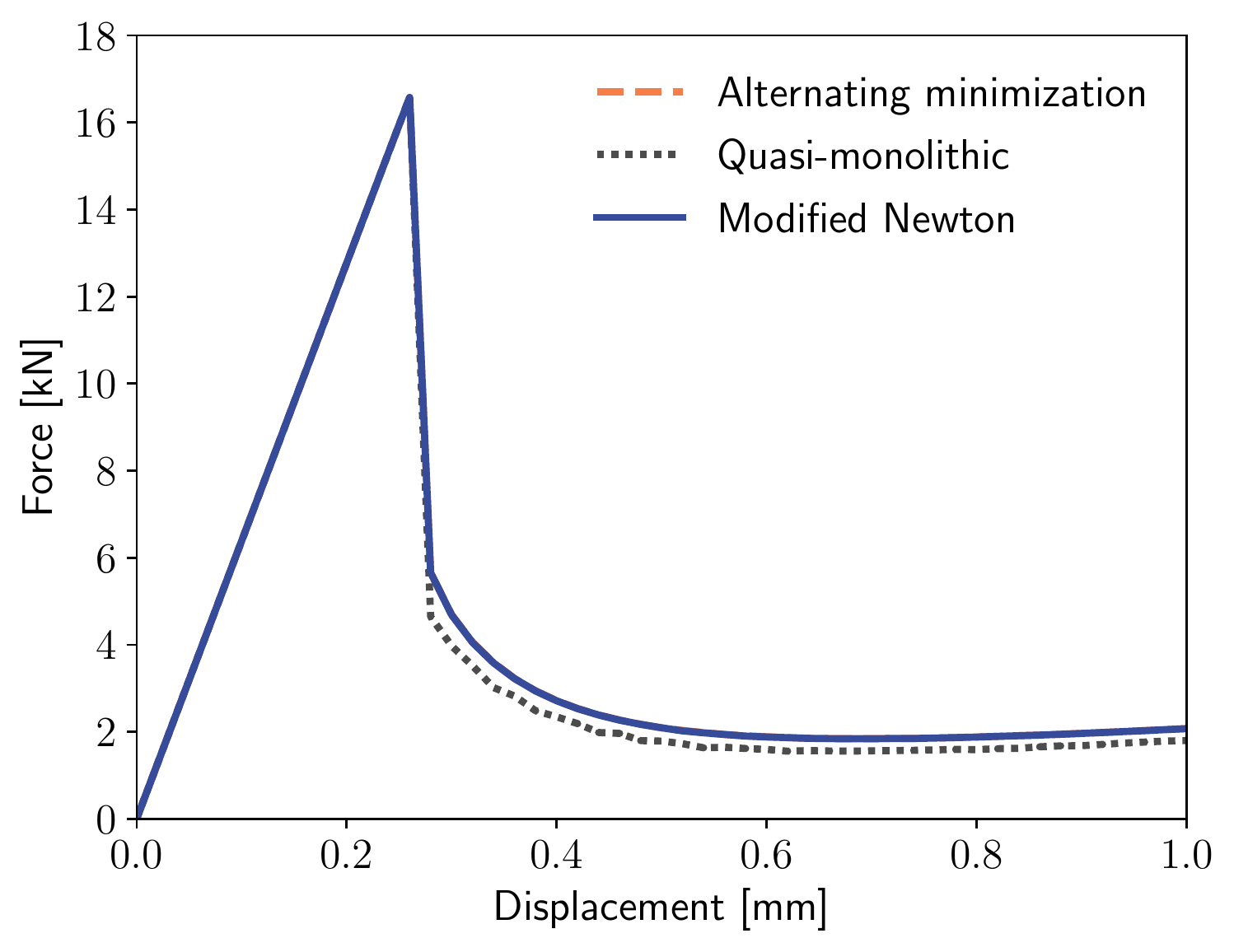}
	\caption{Load-displacement curves resulting from all three methods on the L-shaped panel test. The reaction obtained with the staggered and modified Newton solvers are identical. The response obtained with the quasi-monolithic scheme does not agree with the solutions from the two other methods.}
	\label{fig:LShapedFD}
\end{figure}

The crack paths obtained with the three solvers for the L-shaped panel tests are shown in Figure~\ref{fig:CrackLShaped}.
The crack patterns resulting from the alternating minimization and the modified Newton solvers are identical and in agreement with results found in the literature, e.g.~\cite{wick_error-oriented_2017,ambati_review_2015,mesgarnejad_validation_2015}, where the crack first propagate with an angle and stabilizes in the horizontal direction. A small deviation is observed at the end of the crack path but the results still concur with experimentally observed fractures~\cite{winkler_traglastuntersuchungen_2001}. Figure~\ref{fig:LShapedFD} shows the load-displacement curves obtained for the L-shaped panel with the three algorithms. Again here, the alternating minimization and modified Newton solvers are in agreement. The response also concurs with the results obtained without the research for a global minimum in~\cite{mesgarnejad_validation_2015}. The difference in peak load can be explained by the absence of a tension/compression split and the crack initiation performed in~\cite{mesgarnejad_validation_2015}.

The crack and the load-displacement curve obtained with the quasi-monolithic scheme are in agreement with neither the numerical results obtained from the two other methods nor the experimental results reported in~\cite{winkler_traglastuntersuchungen_2001}. Furthermore, the obtained crack path is physically questionable since the initiation angle does not agree with the maximum hoop stress criterion, stating that the crack should propagate in the direction maximizing the opening stress~\cite{erdogan_crack_1963}. This error can be attributed to the extrapolation correction loop, which in the present case appears to have significantly influenced the solution. However, either no convergence to the implicit solution or no convergence to a stationary point was obtained with different tolerances $\mathtt{TOL}_{QM}$ or different uniform time refinements.

\begin{figure}[p!]
	\centering
	\includegraphics[width=0.63\textwidth]{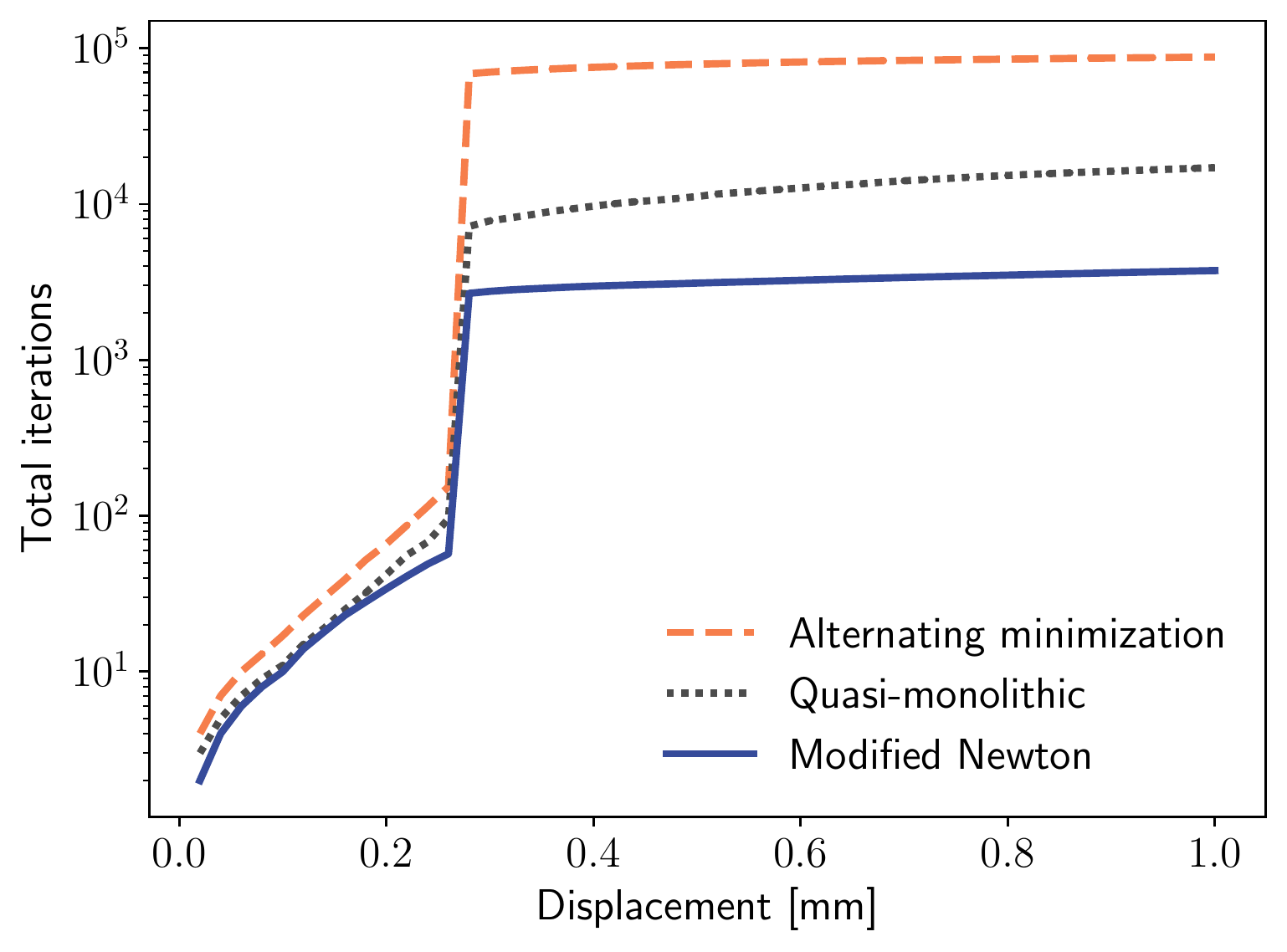}
	\caption{Cumulative number of iterations required by the different methods to solve the L-shaped panel test. The modified Newton solver requires more than 10 times less iterations than the alternating method. The iterations reported for the quasi-monolithic scheme should be interpreted carefully since its solution did not converge to the solution of the other two methods.}
	\label{fig:LShapedIT}
\end{figure}

Figure~\ref{fig:LShapedIT} presents the performance of the three algorithms for the L-shaped panel test, although the results for the extrapolated scheme should be interpreted carefully since the method did not converge to the same solution. As can be seen, the final total iterations required by the modified Newton solver is lower than the alternating minimization algorithm by more than 10 times. Table~\ref{table:LShaped} summarizes the performances of the three methods for the L-shaped panel test. As noted in Table~\ref{table:LShaped}, a staggering number of iterations was required by the alternating minimization algorithm to solve increment 14 ($\bar{u}_y = 0.28$ mm). In comparison, the modified Newton solver required 26 times less iterations for the same increment. \hl{Nevertheless, the 2618 iterations of the modified Newton solver for this time step remain large. Additionally, 2113 iterations required inertia corrections, representing approximately $19\%$ of the total computation time. The high computational cost for the three solvers can be explained by the sudden propagation of a large crack in increment 14 and possibly ill-conditioned systems. However, the conditioning of the systems was not further investigated. Regardless, an important acceleration of the computation time by a factor of $\approx 11.9$ was still observed with the modified Newton solver.}

\begin{table}[p!]
\centering
\caption{Computational performances of the methods investigated on the L-shaped panel benchmark. \hl{For the modified Newton method, the number of iterations requiring inertia correction (IC) and the total time spent computing the inertia corrections are indicated.}}
\label{table:LShaped}
\addtolength{\leftskip} {-2cm}
\addtolength{\rightskip}{-2cm}
\setlength{\tabcolsep}{6pt}
\renewcommand{\arraystretch}{1.2}
\begin{tabular}{cccccc}
\hline
Method                   & Max it. / inc.  & Total it.  & IC it.  & Total time [s] & IC time [s]           \\ \hline
Alternating minimization & 68442            & 87706  & -   & 198377            & -             \\
Quasi-monolithic         & 7093             & 17117   & -   & 62697             & -             \\
Modified Newton          & 2618             & 3743    & \hl{2113}  & 16668             & \hl{3090}             \\ \hline
\end{tabular}
\end{table}

\subsection{Discussion of the results}

We recall that the alternating minimization algorithm, the quasi-monolithic scheme and the modified Newton method all used the same numerical implementation and treatment of the irreversibility condition. The difference remained in the algorithm used to obtain a stationary point or local minimum. The benchmarking on the four tests, selected from the phase-field literature, showed that both the extrapolated scheme and the modified Newton method offered an increase in computational efficiency, when compared to the staggered algorithm. However, once damage started propagating, the semi-implicit extrapolated scheme required additional correction to converge to the implicit solution. Consequently, the modified Newton method was found to be, by far, the most efficient solver. For example, on the SENP-shear test, the modified Newton solver required up to 11 and 3 times less iterations by increment than the alternating minimization and quasi-monolithic algorithms, respectively. For the L-shaped panel test, the increase in efficiency obtained with the modified Newton solver implied a decrease in computation time by a factor of 12 over the staggered solver.

The alternating minimization algorithm remains a powerful and versatile method since it can be applied to non-variational models and implemented using a variational inequality solver, like reported in~\cite{marigo_overview_2016,tanne_crack_2018}. The performance of the extrapolated scheme can probably be enhanced with locally refined time steps and a well-tuned extrapolation. It can also be applied to non-variational formulation. \hl{Nevertheless}, convergence problems were encountered with the quasi-monolithic scheme, as noted for the L-shaped panel test. It is possible that the preconditioned GMRES solver reported in~\cite{wick_modified_2017} would improve the method. The numerical parameters appearing in the inertia correction algorithm of the modified Newton method could also be tuned. \hl{However, it was shown for the first three benchmarks that only very few iterations required inertia correction and that the calculation of the correction represented at most 4\% of the total computational cost. Furthermore, the results of the four tests clearly exhibit that, when using a variational formulation and enforcing the irreversibility condition with a quadratic penalty, the modified Newton solver significantly outperforms the alternating minimization and quasi-monolithic algorithms for quasi-static brittle fracture without requiring any tuning.}

Finally, as already mentioned, convergence issues were sporadically observed with the quasi-monolithic scheme and a surprisingly high number of iterations were necessary for the three solvers on the L-shaped panel test. The latter issues are symptomatic of ill-conditioned non-linear systems and could be attributed to the quadratic penalty term used to enforce the irreversibility condition, even if the optimal penalty parameter from \cite{gerasimov_penalization_2019} was used. As presented in Section \ref{irreversibility}, other methods of enforcing the irreversibility condition are available in the literature. Now that an implicit monolithic solver has been identified, the benchmarking of the schemes used to enforce the $\dot{d} \geq 0$ constraint could allow the identification of a method reducing the issues associated to bad conditioning and minimizing the computational cost associated to the solution of phase-field models.

\section{Conclusion} \label{Conclusions}

A robust, efficient and straightforward solver is still lacking the variational phase-field fracture literature. In an attempt to fill this gap, we proposed a fully monolithic solver based on a modified Newton method with inertia correction and an energy line-search. Additionally, we augmented the quasi-monolithic algorithm with an extrapolation correction loop controlled by a damage-based criterion. Although not perfect, the modified quasi-monolithic scheme allowed us to compare its performance with the alternating algorithm and the modified Newton solver. Through the four test cases presented involving different modes of propagation and initiation, we showed the modified Newton solver to be robust and more efficient than the alternating minimization algorithm and the quasi-monolithic scheme. More precisely, the proposed monolithic solver was shown to be up to $\approx 12$ and $\approx 6$ times faster than the staggered and extrapolated schemes. Furthermore, the performance of the monolithic method did not appear to be problem-dependent, requiring no parameter adjustment, and showing consistent efficiency gains across the test cases.

Additional work can still be done to enhance the performance of phase-field solvers. A study concerning the identification of numerical parameters maximizing the efficiency of the modified Newton solver could be performed. As suggested in the discussion, the benchmarking of the methods used to enforce the irreversibility condition could also help further reduce the computational cost of the phase-field method. However, the combination of the modified Newton solver with an adaptive mesh technique could already facilitate the application of phase-field models to more complex structure and with sophisticated constitutive laws.

\section*{Acknowledgements}
We acknowledge the support of the Natural Sciences and Engineering Research Council of Canada (NSERC) and the Fonds de recherche du Québec - Nature et technologies (FRQNT).

\appendix
\section{Comparison of the original and corrected quasi-monolithic schemes} \label{Appendix}

\begin{figure}[p!]
	\centering
	\includegraphics[width=0.65\textwidth]{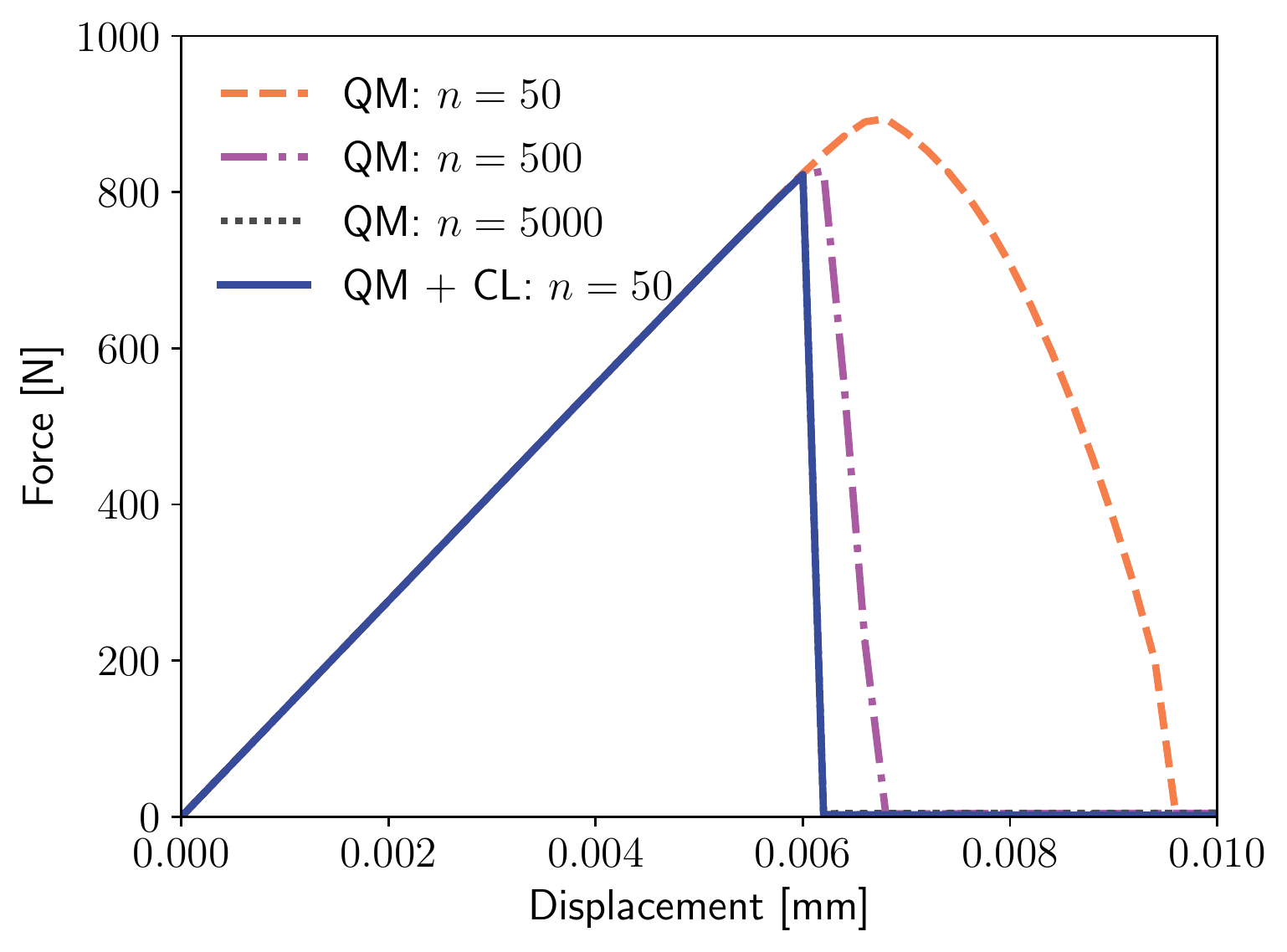}
	\caption{\hl{Load-displacement curves obtained with the original quasi-monolithic scheme (QM) for different time step refinements and with the quasi-monolithic scheme augmented with the extrapolation correction loop (QM+CL) on the SENP-tension test. The solutions produced by the original quasi-monolithic algorithm slowly converge to the expected brutal crack propagation with $n$ increasing. The scheme relying on the extrapolation correction loop captures the brutal crack propagation using only 50 time steps.}}
	\label{fig:Tensile_QM_FD}
\end{figure}

\hl{The original quasi-monolithic scheme~\mbox{\cite{heister_primal-dual_2015}}, as described in Section~\mbox{\ref{Quasi-monolithic scheme}}, and the proposed quasi-monolithic scheme with an extrapolation correction loop, as described in Algorithm~\mbox{\ref{Quasi-mono}}, were applied to the SENP-tensile test using the same geometry, material properties, and numerical parameters as in Section~\mbox{\ref{SENP-tensile}}. Figure~\mbox{\ref{fig:Tensile_QM_FD}} compares the force-displacement solutions obtained with the two quasi-monolithic algorithms. As can be seen in Figure~\mbox{\ref{fig:Tensile_QM_FD}}, the original quasi-monolithic algorithm is sensitive to the step size. For example, with $n = 50$, the force-displacement solution is delayed, with the softening associated to crack propagation happening later and slower. When increasing the number of time steps, with $n = 500$, the softening becomes more brutal, converging to the expected sharp solution. With $n = 5000$, the crack completely propagates within a very small applied displacement, as expected for the SENP-tensile test. At the opposite, the proposed quasi-monolithic scheme with an extrapolation correction loop captures the same brutal propagation using only 50 time steps.}

\begin{figure}[p!]
	\centering
	\includegraphics[width=0.65\textwidth]{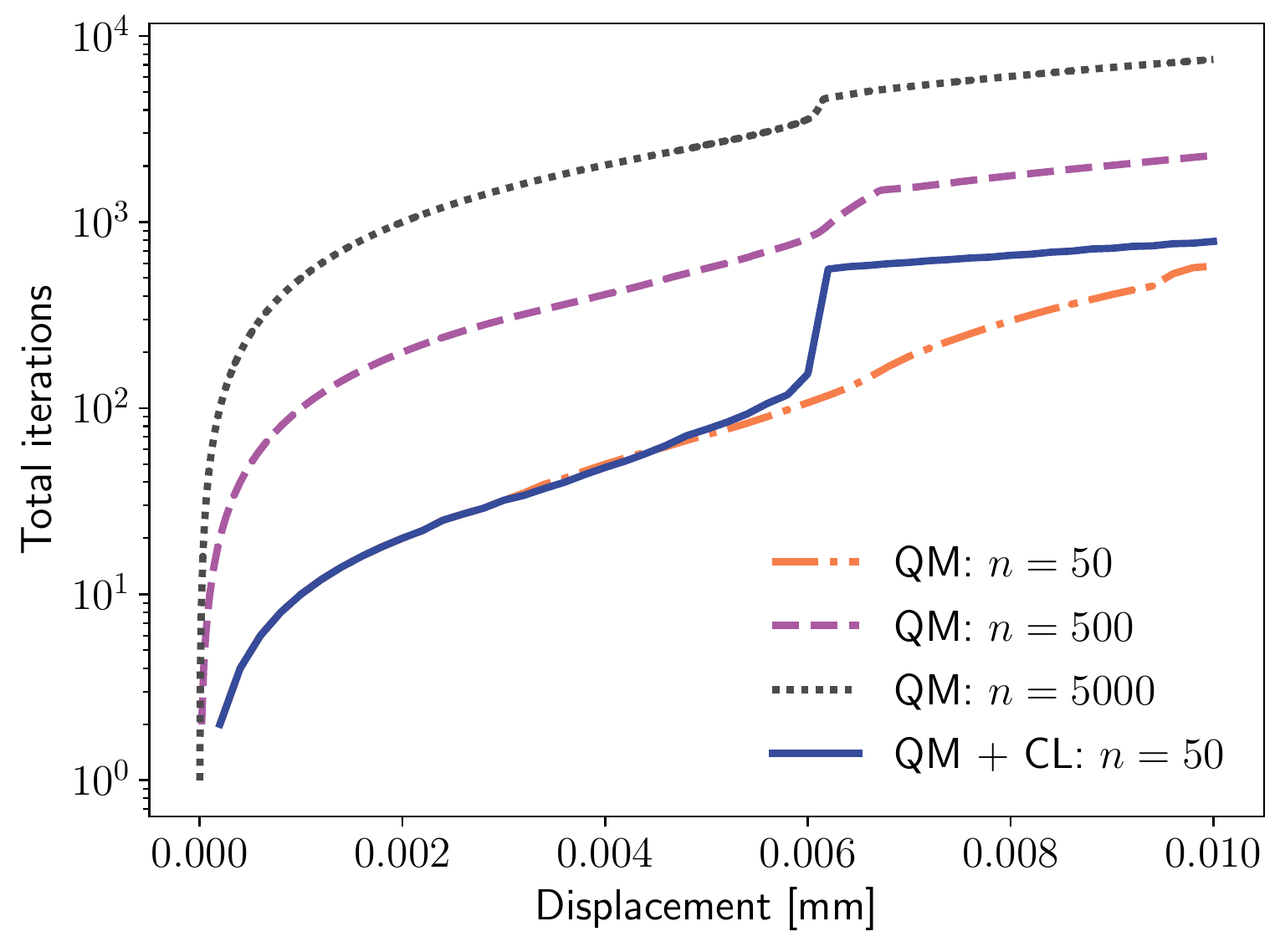}
	\caption{\hl{Total number of iterations required by the original quasi-monolithic scheme (QM) for different time step refinements and the quasi-monolithic scheme augmented with the extrapolation correction loop (QM+CL) to converge on the SENP-tension test. The time step refinements imply additional iterations for the original quasi-monolithic scheme. The extrapolation correction loop requires many iterations for the increment with unstable crack propagation, but remains overall more efficient since only 50 time increments are used.}}
	\label{fig:Tensile_QM_IT}
\end{figure}

\hl{Figure~\mbox{\ref{fig:Tensile_QM_IT}} presents the total number of iterations required by the original extrapolated scheme, with different uniformly refined time steps, and the proposed scheme with the extrapolation correction loop. As shown in Figure~\mbox{\ref{fig:Tensile_QM_IT}}, the additional increments before and after the propagation imply a significant increase in computational cost for the original extrapolated scheme. With $n = 500$, the solution is less accurate while being computationally more expansive than the solution obtained with the proposed extrapolated scheme using only 50 time steps. Therefore, although the uniformly refined time steps allow the original quasi-monolithic scheme to eventually capture the brutal propagation, the proposed extrapolated scheme with the correction loop captures the time converged solution much more efficiently.}

\bibliography{mybibfile}

\end{document}